\documentclass[11pt]{article}
\usepackage{graphicx,ifthen,color,algorithm,palatino}
\usepackage{amsmath,amsthm,amssymb,amsfonts,verbatim}
\usepackage{mathrsfs,accents,setspace}
\newboolean{DoubleSpaced}
\setboolean{DoubleSpaced}{false}
\def\myand{\&\ }
\def\Psc{{\mathcal P}}
\def\ba{\boldsymbol{a}}
\def\bp{\boldsymbol{p}}
\def\real{{\mathbb R}}
\def\by{\boldsymbol{y}}
\def\thickarrow{\longleftarrow}
\def\bTheta{\boldsymbol{\Theta}}
\def\bone{\boldsymbol{1}}
\def\bomega{\boldsymbol{\omega}}
\def\biota{\boldsymbol{\iota}}
\def\smhalf{{\textstyle{\frac{1}{2}}}}
\def\bb{\boldsymbol{b}}
\def\bbeta{\boldsymbol{\beta}}
\def\bu{\boldsymbol{u}}
\def\simind{\stackrel{{\tiny \mbox{ind.}}}{\sim}}
\def\expit{\mbox{expit}}
\def\bzero{\boldsymbol{0}}
\def\bI{\boldsymbol{I}}
\def\btheta{\boldsymbol{\theta}}
\def\lboxit#1{\vbox{\hrule\hbox{\vrule\kern6pt
      \vbox{\kern6pt#1\kern6pt}\kern6pt\vrule}\hrule}}
\def\bell{\boldsymbol{\ell}}
\def\bX{\boldsymbol{X}}
\def\bmu{\boldsymbol{\mu}}
\def\bSigma{\boldsymbol{\Sigma}}
\def\bOmega{\boldsymbol{\Omega}}
\def\bU{\boldsymbol{U}}
\def\diag{\mbox{diag}}
\def\bd{\boldsymbol{d}}
\def\bz{\boldsymbol{z}}
\def\naturalNumbers{{\mathbb N}}
\def\bsigma{\boldsymbol{\sigma}}
\def\bxnew{\bx_{\mbox{{\tiny new}}}}
\def\bx{\boldsymbol{x}}
\def\bdeta{\boldsymbol{\eta}}
\def\sigeps{\sigma_{\varepsilon}}
\def\bZ{\boldsymbol{Z}}
\def\bC{\boldsymbol{C}}
\def\bc{\boldsymbol{c}}
\def\bB{\boldsymbol{B}}
\def\bib{\vskip12pt\par\noindent\hangindent=1 true cm\hangafter=1}
\def\Var{\mbox{Var}}
\def\infint{\int_{-\infty}^{\infty}}
\def\hhat{{\widehat h}}
\def\bUOmega{\bU_{\mbox{\tiny{$\bOmega$}}}}
\def\bdOmega{\bd_{\mbox{\tiny{$\bOmega$}}}}
\def\bxcurr{\bx_{\tiny\mbox{curr}}}
\def\bycurr{\by_{\tiny\mbox{curr}}}
\def\bCcurr{\bC_{\tiny\mbox{curr}}}
\def\bXcurr{\bX_{\tiny\mbox{curr}}}
\def\betazTrue{\beta_{0,\mbox{{\tiny true}}}}
\def\betaoTrue{\beta_{1,\mbox{{\tiny true}}}}
\def\fTrue{f_{\mbox{{\tiny true}}}}
\def\xnew{x_{\mbox{{\tiny new}}}}
\def\ynew{y_{\mbox{{\tiny new}}}}

\def\xOneNew{x_{1\mbox{{\tiny new}}}}
\def\xTwoNew{x_{2\mbox{{\tiny new}}}}

\def\Nburn{N_{\mbox{\tiny burn}}}
\def\Nkept{N_{\mbox{\tiny kept}}}
\def\pDens{\mathfrak{p}}
\def\oneseventh{{\textstyle{1\over7}}}
\def\twosevenths{{\textstyle{2\over7}}}
\def\foursevenths{{\textstyle{4\over7}}}
\def\sixsevenths{{\textstyle{6\over7}}}
\def\SystematicResample{\textsc{\footnotesize SystematicResample}}
\def\SystematicResampleSecHead{\textsc{SystematicResample}}
\def\npartic{M}
\def\iStt{i_{\mbox{\tiny stt}}}
\def\iEnd{i_{\mbox{\tiny end}}}

\def\ssigsq{s_{\sigma^2}}
\def\bomegaSMC{\bomega_{\mbox{\tiny SMC}}}
\def\bomegaRW{\bomega_{\mbox{\tiny RW}}}
\def\bbetaSMC{\bbeta_{\mbox{\tiny SMC}}}
\def\bbetaRW{\bbeta_{\mbox{\tiny RW}}}
\def\bdetaSMC{\bdeta_{\tiny\mbox{SMC}}}
\def\bdetaRW{\bdeta_{\tiny\mbox{RW}}}
\def\bbetaSMCm{\bbeta_{\mbox{\tiny SMC},m}}
\def\bbetaZeroSMC{\bbeta_{0\mbox{\tiny SMC}}}
\def\bbetaOneSMC{\bbeta_{1\mbox{\tiny SMC}}}
\def\bbetaTwoSMC{\bbeta_{2\mbox{\tiny SMC}}}
\def\buSMC{\bu_{\mbox{\tiny SMC}}}
\def\buRW{\bu_{\mbox{\tiny RW}}}
\def\buOneSMC{\bu_{1\mbox{\tiny SMC}}}
\def\buTwoSMC{\bu_{2\mbox{\tiny SMC}}}
\def\sigsqSMC{\bsigma^2_{\mbox{\tiny SMC}}}
\def\sigsqEpsSMC{\bsigma^2_{\varepsilon\mbox{\tiny SMC}}}
\def\aEpsSMC{a_{\varepsilon\mbox{\tiny SMC}}}
\def\aSMC{\ba_{\mbox{\tiny SMC}}}
\def\bsigsqUSMC{\bsigma^2_{u\mbox{\tiny SMC}}}
\def\baUSMC{\ba_{u\mbox{\tiny SMC}}}
\def\logw{\bell}
\def\hFP{h_{\mbox{\tiny FP}}}
\def\hKDE{h_{\mbox{\tiny KDE}}}
\def\hhatFP{\hhat_{\mbox{\tiny FP}}}
\def\hhatKDE{\hhat_{\mbox{\tiny KDE}}}

\def\bthetaSMC{\btheta_{\mbox{\tiny SMC}}}
\def\LMparmsSMC{\left[
\begin{array}{c}
\bbetaSMC\\[0ex]
\aSMC\\[0ex]
\sigsqSMC
\end{array}
\right]}
\def\LMMparmsSMC{\left[
\begin{array}{c}
\bbetaSMC\\[0ex]
\buSMC\\[0ex]
\sigsqEpsSMC\\[0ex]
\aEpsSMC\\[0ex]
\bsigsqUSMC\\[0ex]
\baUSMC
\end{array}
\right]}
\def\bbetabuSMC{\left[
\begin{array}{c}
\bbetaSMC\\
\buSMC
\end{array}
\right]
}
\def\bbetabuRW{\left[
\begin{array}{c}
\bbetaRW\\
\buRW
\end{array}
\right]
}
\def\LMparmsSMC{\left[
\begin{array}{c}
\bbetaSMC\\[0ex]
\aSMC\\[0ex]
\sigsqSMC
\end{array}
\right]}
\def\GLMMparmsSMC{\left[
\begin{array}{c}
\bbetaSMC\\[0ex]
\buSMC\\[0ex]
\bsigsqUSMC\\[0ex]
\baUSMC
\end{array}
\right]}
\def\bbetabuSMC{\left[
\begin{array}{c}
\bbetaSMC\\
\buSMC
\end{array}
\right]
}
\def\bbetabu{\left[
\begin{array}{c}
\bbeta\\
\bu
\end{array}
\right]
}
\def\batheta{\ba_{\theta}}
\def\bptheta{\bp_{\theta}}
\def\yTy{\textbf{yTy}}
\def\XTy{\textbf{XTy}}
\def\CTy{\textbf{CTy}}
\def\XTX{\textbf{XTX}}
\def\CTC{\textbf{CTC}}
\def\nwarm{n_{\mbox{\tiny warm}}}
\def\nvalid{n_{\mbox{\tiny valid}}}
\def\bywarm{\by_{\mbox{\tiny warm}}}
\def\bXwarm{\bX_{\mbox{\tiny warm}}}
\def\bcnew{\bc_{\mbox{\tiny new}}}

\def\bCwarm{\bC_{\mbox{\tiny warm}}}

\def\aeps{a_{\varepsilon}}
\def\seps{s_{\varepsilon}}

\def\sur{s_{ur}}
\definecolor{ruppertgreen}{rgb}{0,.7,.25}
\definecolor{SteelBlue}{rgb}{0.275,0.51,0.706}
\definecolor{DarkOrange}{rgb}{1,0.549,0}

\begin{document}

\ifthenelse{\boolean{DoubleSpaced}}{\setstretch{1.5}}{}

\vskip5mm
\centerline{\Large\bf Online Semiparametric Regression via}
\vskip1.5pt
\centerline{\Large\bf Sequential Monte Carlo}
\vskip4mm
\centerline{\normalsize\sc Marianne Menictas$\null^1$, Chris J. Oates$\null^2$ 
\myand Matt P. Wand$\null^3$}
\vskip4mm
\centerline{\textit{$\null^1$Grubhub Inc., U.S.A.,
$\null^2$Newcastle University, U.K.}}
\vskip0.5pt
\centerline{\textit{and $\null^3$University of Technology Sydney, Australia}}
\vskip3mm
\centerline{27th August, 2024}
\vskip6mm
\centerline{\large\bf Abstract}
\vskip2mm

We develop and describe online algorithms for performing online semiparametric 
regression analyses. Earlier work on this topic is in Luts, Broderick \myand Wand 
(\textit{J. Comput. Graph. Statist.}, 2014) where online mean field variational Bayes 
was employed. In this article we instead develop sequential Monte Carlo approaches 
to circumvent well-known inaccuracies inherent in variational approaches. 
For Gaussian response semiparametric regression models our new algorithms 
share the online mean field variational 
Bayes property of only requiring updating and storage 
of sufficient statistics quantities of streaming data. In the non-Gaussian case 
accurate online semiparametric regression requires the full data to be kept 
in storage. The new algorithms allow for new options concerning accuracy/speed
trade-offs for online semiparametric regression.


\section{Introduction}\label{sec:intro}

Online semiparametric regression is concerned with rapid online fitting of flexible regression
models, such as generalized additive models, and continuously updated inference as data stream in.
Luts, Broderick \myand Wand (2014) laid out a framework for online, also known as real-time, 
semiparametric regression in the wake of developments over the preceding two decades such as 
Bayesian mixed model-based penalized splines and mean field variational Bayes. 
The setting and motivations are identical
to those of Luts \textit{et al.} (2014) and background material in the earlier article
on the essence of online semiparametric regression also applies here. The crux of 
this article is provision of sequential Monte Carlo alternatives to the online mean field 
variational Bayes approach of Luts \textit{et al.} (2014). 

Online fitting of statistical models for sequentially arriving data 
has a very long history and large literature. The essential goal is that of
obtaining fits and corresponding inference with online updating -- such
that the online results are similar to the batch results. 
Clearly online fitting is preferable in applications in which 
the data arrive rapidly, and repeated batch fitting 
is not computationally feasible. Three recent examples of such situations 
are online fitting of regression models for streaming data such as electronic health 
records and mobile health data (Luo \myand Song, 2023),
online anomaly detection in streaming temporal data (Talagala \textit{et al.}, 2020)
and real-time fitting of item response theory models for ratings of movies
(Weng \myand Coad, 2018).

For semiparametric regression and related areas there is also a large literature
on online fitting, with early contributions such as recursive kernel density 
estimation (e.g. Yamato, 1971; Carroll, 1976) and kernel regression 
(e.g. Krzyzak \myand Pawlak, 1982; Yin \myand Yin, 1996). However, none of
these 20th Century contributions addressed the problem of online smoothing 
parameter choice and, instead, were concerned with the theoretical properties
of kernel estimators for deterministic smoothing parameter sequences.
Bayesian computing developments since the 1990s have given
rise to online semiparametric regression schemes for which the smoothing
parameters are updated in a principled and practical manner. For the 
kriging approach to nonparametric regression, Gramacy \myand Polson (2011)
achieve this via sequential Monte Carlo. As mentioned earlier,
Luts \textit{et al.} (2014) achieved it via mean field variational Bayes.
The essence of the present article is sequential Monte Carlo methodology for 
online semiparametric regression according to the mixed model-based splines 
approach advocated in the Ruppert, Wand \myand Carroll (2003)
monograph. This approach has the attraction of only requiring the mixed model extension 
of ordinary linear models to achieve nonparametric and semiparametric regression fitting 
and inference.

The algorithms of Luts \textit{et al.} (2014) have the attractiveness of being \emph{purely}
online in that, when a new vector of observations arrives, the approximate Bayesian semiparametric
regression fit is updated \emph{without having to store or access previous observations}.
Instead, only key sufficient statistics need to be updated -- after which the new observation
vector can be discarded. A disadvantage of Luts \textit{et al.} (2014) is that the inference
is subject to varying degrees of inaccuracy due to mean field-type variational approximation
error. This is particularly the case for regression models with non-Gaussian responses.

The sequential Monte Carlo-based approach used here is devoid of variational approximations
and produces accurate online semiparametric regression fitting and inference. In the case
of regression models with Gaussian response, purely online fitting and inference is achievable.
However, for regression models with non-Gaussian responses the purely online feature has to 
be sacrificed to overcome the accuracy shortcomings of the Luts \textit{et al.} (2014) approach
and the full data must be kept in storage. The upshot is that this article's online semiparametric
regression approach is more accurate, but not as fast, as the approach used in Luts \textit{et al.} (2014).
Depending on the speed requirements of the application and volume of data requiring storage, the
new sequential Monte Carlo approaches to online semiparametric regression may be preferable.
In short, the contributions of this article provide users with speed/accuracy trade-off options
for online semiparametric regression.

Sequential Monte Carlo methodology of the type used here originates with Kong \textit{et al.} (1994).
Lie \myand Chen (1998) applied the approach to dynamical systems and introduced the 
\emph{sequential Monte Carlo} idiom. Other key early contributions 
include Gilks \myand Berzuini (2001) and  Pitt \myand Shephard (1999). A general theoretical 
framework for sequential Monte Carlo was devised by Del Moral \textit{et al.} (2006). 
A comprehensive and contemporary overview of sequential Monte Carlo is provided by
Chopin \myand Papaspiliopoulos (2020).

Section \ref{sec:prelimInfra} lays down some preliminary infrastructure that is intrinsic
to the sequential Monte Carlo approach to online semiparametric regression. In 
Section \ref{sec:GaussResp} we treat Gaussian response models, starting with multiple
linear regression. For this special case  the new methodology is relatively simple and 
the essence of the general approach can be elucidated in a reasonably concise manner.
Section \ref{sec:generResp} then tackles the more challenging non-Gaussian response situation.
In Section \ref{sec:illus} we present some illustrations of that demonstrate good inferential 
accuracy of online sequential Monte Carlo and contrast it with the patchy performance
of online mean field variational Bayes. Some concluding remarks are made in Section \ref{sec:conclud}.

\section{Preliminary Infrastructure}\label{sec:prelimInfra}

Online semiparametric regression via sequential Monte Carlo depends on 
some fundamental concepts and results, which we lay out in this section.
Throughout this section $I(\Psc)$ denotes the indicator of the proposition 
$\Psc$ being true.

\subsection{Discrete Distribution Nomenclature}

Suppose that a discrete random variable assumes the 
values of $5$, $11$ and $13$ with probabilities $\twosevenths$,
$\foursevenths$ and $\oneseventh$ respectively. Its 
\emph{probability mass function}, $\pDens$, is:
\begin{equation}
\pDens(5)=\twosevenths,\ \ \quad \pDens(11)=\foursevenths,\ \ \quad \pDens(13)=\oneseventh
\quad\mbox{and}\quad \pDens(x)=0\ \mbox{if $x\notin\{5,11,13\}$}.
\label{eq:KangarooPoint}
\end{equation}
We say that $\pDens$ has \emph{atoms} $\ba=(5,11,13)$
and \emph{probabilities} $\bp=(\twosevenths,\foursevenths,\oneseventh)$.
The corresponding \emph{cumulative distribution function} is
$$F(x;\ba,\bp)=\twosevenths I(x\le5)+\foursevenths I(x\le11)+\oneseventh I(x\le13),\ 
x\in\real
$$
and \emph{quantile function} is
\begin{equation}
Q(q;\ba,\bp)\equiv\inf\{x\in\real:q\le F(x)\}=5+6I(q>\twosevenths)+2I(q>\sixsevenths),\ 0\le q\le1.
\label{eq:quantileDiscrete}
\end{equation}
The concepts illustrated here are, of course, very basic and straightforwardly extended
to general discrete random variables.

\subsection{Discrete Posterior Distribution Approximations}\label{sec:discAppx}

Let $\theta$ be a generic parameter in a Bayesian statistical model that takes values over a continuum
such as $\real$, $\real_+$ or $[0,1]$. Also, let $\bycurr$ denote the currently observed data. 
Then, within the Bayesian model, $\theta$ is a continuous 
random variable and its posterior distribution is characterized by the probability density 
function $\pDens(\theta|\bycurr)$. An intrinsic feature of online semiparametric regression 
via sequential Monte Carlo is sequential approximation of $\pDens(\theta|\bycurr)$ by 
\emph{probability mass functions} as new observations arrive. To repeat: even though
$\pDens(\theta|\bycurr)$ is a probability density function, it is sequentially approximated
by probability mass functions as the data stream in.

Suppose that a new observation $\ynew$ has just been read in. The currently observed data is then
updated according to $\bycurr\thickarrow(\bycurr,\ynew)$.
The current posterior density function of $\theta$, $\pDens(\theta|\bycurr)$, 
is updated to be a probability mass function having atoms $\batheta$ and probabilities $\bptheta$.
Then the current posterior mean of $\theta$ is approximated by $(\bptheta)^T\batheta$ and, for
example, a current approximate 95\% credible interval for $\theta$ is 
$$\big(Q(0.025;\batheta,\bptheta),Q(0.975;\batheta,\bptheta)\big)$$
where $Q(\cdot;\batheta,\bptheta)$ is the quantile function corresponding to the probability
mass function having atoms $\batheta$ and probabilities $\bptheta$. 

\subsection{The \SystematicResampleSecHead\ Algorithm}

A fundamental component of sequential Monte Carlo procedures is that of drawing
a sample from a $d$-variate discrete distribution having $M$ atoms. The sample
size is also $M$. Drawing a simple random sample is usually called 
\emph{multinomial resampling} in the sequential Monte Carlo literature.
However, in their Section 9.7, Chopin \myand Papaspiliopoulos (2020)
advise against multinomial sampling due to its poor performance 
compared with other schemes. A simple alternative scheme is
\emph{systematic resampling}, which is the one that we adopt here. 
The operational steps are provided by 
the \SystematicResample\ algorithm, listed as Algorithm \ref{alg:SystematicSample},
which involves storing the atoms as columns of a $d\times M$ matrix.

%
%
\begin{algorithm}[h]
\begin{center}
\begin{minipage}[h]{145mm}
\begin{small}
\vskip1mm
\begin{itemize}
\setlength\itemsep{4pt}
\item[] Inputs: $\bTheta\ (d\times M)$, $\bp\ (M\times1)$ such that 
all entries of $\bp$ are non-negative and $\bp^T\bone=1$ 
%
%
%
%
%
%
\item[] $\bomega_1\thickarrow\mbox{the $M\times1$ vector of cumulative sums of the entries of $\bp$}$
\ \ \ ;\ \ \ $\bomega_2\thickarrow M\bomega_1$
\item[]$u\thickarrow\mbox{draw from the Uniform$(0,1)$ distribution}$
\ \ \ ;\ \ \ $\omega_3\thickarrow u$\ \ \ ;\ \ \ $k\thickarrow1$
\item[] for $m = 1,\ldots M$
\begin{itemize}
\item[] while \big\{$\omega_3<(\bomega_2)_k\big\}\ k\thickarrow k + 1$
\ \ \ ;\ \ \ $(\biota)_m\thickarrow k$\ \ \ ;\ \ \ $\omega_3\thickarrow\omega_3+1$
\end{itemize}
\item[] $\bTheta\thickarrow\mbox{$d\times M$ matrix with the current columns
of $\bTheta$ replaced by those indexed by $\biota$}$
%
\item[] Output: $\bTheta\ (d\times M)$
\end{itemize}
\end{small}
\end{minipage}
\end{center}
\caption{\textit{The} \SystematicResample\ \textit{algorithm.}}
\label{alg:SystematicSample} 
\end{algorithm}
%
%

\vfill\eject

An example of the last step of \SystematicResample\ is as follows:
if $d=3$, $M=5$ and $\biota=(3,3,5,2,2)$ then the inputted matrix
$$
\left[
\begin{array}{rrrrr}
1 & 4 & 7 & 10 & 13\\
2 & 5 & 8 & 11 & 14\\
3 & 6 & 9 & 12 & 15
\end{array}
\right]
\quad\mbox{is outputted as}\quad
\left[
\begin{array}{rrrrr}
7 & 7 & 13 & 4 & 4\\
8 & 8 & 14 & 5 & 5\\
9 & 9 & 15 & 6 & 6
\end{array}
\right].
$$

A justification of Algorithm \ref{alg:SystematicSample} is
given in Section \ref{sec:DDSjustif} of the supplement.

\subsection{Distributional Definitions}\label{sec:DistDefns}

The random variable $x$ has an Inverse Gamma distribution with parameters
$\kappa$ and $\lambda$, written $x\sim\mbox{Inverse-Gamma}(\lambda,\kappa)$,
if and only if its density function is
$$\pDens(x)=\frac{\lambda^{\kappa}}{\Gamma(\kappa)}\,x^{-\kappa-1}\exp(-\lambda/x)I(x>0).$$

In the semiparametric regression models to follow it is usual to place a Half Cauchy prior 
distribution on  each of the standard deviation parameters. Let $\sigma$ denote a typical 
standard deviation parameter. Then the notation $\sigma\sim\mbox{Half-Cauchy}(s)$ means 
that $\sigma$ has density function
$$\pDens(\sigma)=\frac{2I(\sigma>0)}{\pi\,s\{1+(\sigma/s)^2\}}.$$
However, via introduction of an auxiliary random variable $a$, we can express 
$\sigma\sim\mbox{Half-Cauchy}(s)$ as follows:
\begin{equation}
\sigma^2|a\sim\mbox{Inverse-Gamma}(\smhalf,1/a),\quad a\sim\mbox{Inverse-Gamma}(\smhalf,1/s^2).
\label{eq:ShirleyBenuik}
\end{equation}
Representation (\ref{eq:ShirleyBenuik}) has the attraction of leading to draws from
standard distributions in our sequential Monte Carlo schemes.

\subsection{Vector Definitions and Conventions}

The symbol $\bone$ denotes a column vector having all entries equal to $1$.
If $\ba$ is column vector then $\Vert\ba\Vert=\sqrt{\ba^T\ba}$ denotes the
Euclidean norm of $\ba$. If $\ba$ and $\bb$ are both $d\times1$ vectors then $\ba\odot\bb$ denotes 
the $d\times1$ vector containing the element-wise products of 
the entries of $\ba$ and $\bb$. Similarly $\ba/\bb$ is the $d\times1$ vector
of element-wise quotients. Also, if $s$ is a scalar-to-scalar function then 
$s(\ba)$ is the $d\times1$ vector containing the element-wise function evaluations.
An example is $\exp\big([7\ \ 4\ \ 6]^T\big)=\big[\exp(7)\ \exp(4)\ \exp(6)\big]^T$.
Also, $\max(\ba)$ denotes the largest entry in $\ba$.

\subsection{Overview of Online Semiparametric Regression via Sequential Monte Carlo}\label{sec:overview}

\begin{figure}[!h]
\centering
\includegraphics[width=0.65\textwidth]{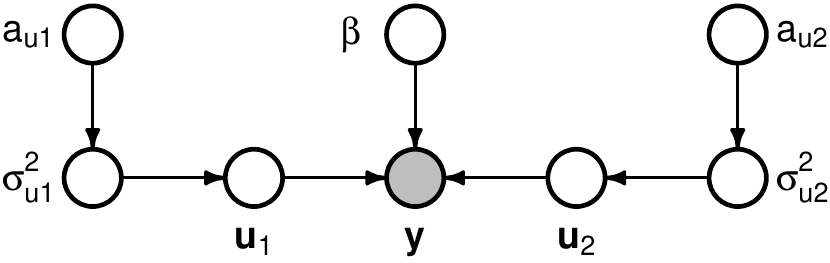}
\caption{\textit{Directed acyclic graph corresponding to the Bayesian logistic
additive model (\ref{eq:lastGasp}). The shading indicates that the $\by$ node  
is observed.}}
\label{fig:logisAddModDAG}
\end{figure}

Loosely speaking, \emph{semiparametric regression} involves extensions of 
parametric linear models in which non-linear effects are handled using 
suitable basis functions and penalization (e.g. Ruppert \textit{et al.}, 2003).
Special cases include nonparametric regression, generalized additive models, varying-coefficient models
and generalized additive mixed models. The version of semiparametric regression which we use here 
is Bayesian models for which the non-linear effects correspond to mixed model-based penalized splines.
As an illustrative example, consider the regression-type data set with responses $y_i\in\{0,1\}$ and 
bivariate continuous predictors $(x_{1i},x_{2i})$, $1\le i\le n$. Then a \emph{logistic additive model} is 
\begin{equation}
{\setlength\arraycolsep{1pt}
\begin{array}{rcl}
&&y_i|\bbeta,\bu_1,\bu_2\simind\mbox{Bernoulli}\left(
\expit\Bigg(
\beta_0+\beta_1x_{1i}+{\displaystyle\sum_{k=1}^{K_1}}u_{1k}z_{1k}(x_{1i})
+\beta_2x_{2i}+{\displaystyle\sum_{k=1}^{K_2}}u_{2k}z_{2k}(x_{2i})\Bigg)\right),\\[3ex]
&&\beta_0,\beta_1,\beta_2\simind N(0,\sigma_{\beta}^2),\quad
\bu_r|\sigma_{ur}^2\simind N(\bzero,\sigma_{ur}^2\bI_{K_r}),\quad 
\sigma_{ur}^2|a_{ur}\simind \mbox{Inverse-Gamma}(\smhalf,1/a_{ur}),\\[2ex]
&&
a_{ur}\simind\mbox{Inverse-Gamma}\big(\smhalf,1/s^2_{\sigma^2}\big),\quad r=1,2,
\end{array}
}
\label{eq:lastGasp}
\end{equation}
where $\simind$ stands for ``independently distributed as'' and 
$\expit(x)\equiv 1/(1+e^{-x})$. The functions 
$\{z_{1k}(\cdot):1\le k\le K_1\}$ and $\{z_{2k}(\cdot):1\le k\le K_2\}$ are
suitable spline bases (see e.g. Wand \myand Ormerod, 2008)
for the $x_1$ and $x_2$ non-linear effects. 
Also, $\sigma_{\beta}>0$ and $s_{\sigma^2}>0$ are user-specified
hyperparameters. Figure \ref{fig:logisAddModDAG} is a directed acyclic graph representation 
of (\ref{eq:lastGasp}) with, for example, $\by$ denoting the vector containing
the $y_i$ data. The $\by$ node is shaded to indicate that it corresponds to
the observed data. Each of the other nodes require inference.

Model (\ref{eq:lastGasp}) and its Figure \ref{fig:logisAddModDAG} representation
exemplifies the approach to Bayesian semiparametric regression used here, with spline
basis function penalization achieved via linear mixed model embedding. Batch fitting
of (\ref{eq:lastGasp}) is straightforward using Bayesian inference engines 
such as \textsf{JAGS} (Plummer, 2022) and \textsf{Stan} (Stan Development Team, 2022) 
(e.g. Harezlak \textit{et al.}, 2018). Our concern here, though, is online 
fitting of (\ref{eq:lastGasp}) as data stream in. Algorithm 5 of Luts \textit{et al.} (2014) 
provides a solution to this problem using online mean field variational Bayes. However, 
this approach is susceptible to poor Bayesian inferential accuracy. Therefore, 
the sequential Monte Carlo alternative is being pursued here.

Typical Bayesian semiparametric regression models have between tens and hundreds of parameters
requiring inference from the response and predictor observations. These include fixed effects,
random effects, spline coefficients and covariance matrix parameters. Let $d$ denote the
total number of such variables and $\btheta$ be the $d\times1$ vector containing them.
For online semiparametric regression, the posterior density function of $\btheta$ 
is sequentially approximated by probability mass functions having $M$ atoms, 
which are referred to as \emph{particles}. The value of $M$ is a user-specified tuning 
parameter and a reasonable default is $M=1000$.

\begin{figure}[!h]

\lboxit{
\begin{center}
\begin{minipage}[h]{140mm}
\null\vskip1mm
\begin{itemize}
\setlength\itemsep{0.2pt}
\item[] Initialise the sample size to be $0$. Initialise key sufficient statistic quantities.
\item[] Initialise the $d\times M$ matrix $\bthetaSMC$ such that each row contains $M$ draws
from the posterior distribution of each component of $\btheta$.
\item[] Initialise the particle probability vector $\bp$ to be the $M\times 1$ vector with
each entry equal to $1/M$.
\item[] Cycle:
\begin{itemize}
\item[] Read in a new observation vector. Increment the sample size by $1$.
\item[] Update key sufficient statistic quantities.
\item[] Use the likelihood of the new observation vector to update $\bp$.
\item[] If the sum of squares of entries of $\bp$ is above a particular threshold then
\begin{itemize}
\item[] Update $\bthetaSMC$ by drawing a sample of size $M$ from the $d$-variate 
discrete distribution with $M$ atoms corresponding to the columns of $\bthetaSMC$
and probability vector $\bp$. This step is facilitated by the  \SystematicResample\ 
algorithm. Set $\bp$ to be the $M\times 1$ vector with each entry equal to $1/M$.
%
\end{itemize}
\item[] Update $\bthetaSMC$ by drawing samples from the current full conditional
distributions of sub-blocks of $\btheta$. Typically, the sub-blocks correspond to 
(1) the coefficients vector and (2) variance or covariance matrix parameters.
\item[] Approximate the current posterior distribution of $\btheta$ by the $d$-variate discrete
distribution with $M$ atoms corresponding to the columns of $\bthetaSMC$ and probability
vector $\bp$. Make inferential summaries of quantities of interest based on the 
current approximate posterior distribution of $\btheta$ as described in 
Section \ref{sec:discAppx}.
\end{itemize}
\item[] until data no longer available or analysis terminated.
\end{itemize}
\end{minipage}
\end{center}
}
\caption{\textit{The sequential Monte Carlo approach to online semiparametric
regression in generic form. Here $\btheta$ is the vector containing all fixed effects,
random effects and covariance matrix parameters in the semiparametric regression
model of interest.}}
\label{fig:genericFormSMC}
\end{figure}

Figure \ref{fig:genericFormSMC} conveys the sequential Monte Carlo approach
to online semiparametric regression in generic terms. Justification for
this scheme, which applies to Bayesian models in general,  
is given in Section \ref{sec:SMCdetails} of the supplement.
Most of the steps in Figure \ref{fig:genericFormSMC}  involve simple calculations. 
The possible exception is the step involving drawing independent samples from the 
current full conditional distributions. For the sub-vectors of $\btheta$
for which the full conditional distribution has a standard form, such as Multivariate
Normal or Inverse Gamma, this step is also straightforward. Moreover, for
such fully Gibbsian settings, the updates only depend on sufficient statistics
of the streaming data such as the sum of squares of the responses. Therefore,
only these sufficient statistics need to be updated and stored for the Gibbsian
situations that arise in Gaussian response semiparametric regression. The 
generalized response situation, with model (\ref{eq:lastGasp}) as an example,
is more challenging due to the current full conditional distributions
having non-standard forms and the need for more elaborate approaches 
that require passes through the current full data.

As explained in Section \ref{sec:fixPartic} of the supplement, the general form of the probability 
vector updates when a new response $\ynew$ and its corresponding predictor data vector, 
arrives is
\begin{equation}
\bp_m^{\tiny\mbox{new}}\propto 
\left\{\frac{\pDens(\btheta|\by_{\tiny\mbox{curr}},\ynew)}
{\pDens(\btheta|\by_{\tiny\mbox{curr}})}\right\}
\bp_m^{\tiny\mbox{curr}},\quad 1\le m\le M.
\label{eq:LeedsBagShop}
\end{equation}
Straightforward algebraic arguments then lead to the updating steps:
\begin{equation}
\begin{array}{rcl}
\bell_m&\longleftarrow&\bell_m+\mbox{log-likelihood of $\ynew$ based on $(\bthetaSMC)_m$},
\quad\ 1\le m\le M,\\[1ex]
\bp^{\tiny\mbox{new}}&\longleftarrow&{\displaystyle\frac{\exp\{\bell-\max(\bell)\}}
{\bone^T\exp\{\bell-\max(\bell)\}}},
\end{array}
\label{eq:ForbesCoffee}
\end{equation}
with logarithms and centring used to mitigate against overflow and underflow in the probability
vector updates. Note that the likelihood of $\ynew$  is 
given by $\pDens(\ynew|\mbox{parents of $\ynew$})$ in the model's directed acyclic graph.
For the illustrative logistic additive model example given by (\ref{eq:lastGasp}) and
Figure \ref{fig:logisAddModDAG} we have 
$$\bell_m\longleftarrow\bell_m+\ynew\eta_m-\log(1+e^{\eta_m}),\quad\ 1\le m\le M,$$
where 
$$\eta_m\equiv (\bbetaZeroSMC)_m
+(\bbetaOneSMC)_{m}\xOneNew+{\displaystyle\sum_{k=1}^{K_1}}(\buOneSMC)_{km}z_{1k}(\xOneNew)
+(\bbetaTwoSMC)_{m}\xTwoNew
+{\displaystyle\sum_{k=1}^{K_2}}(\buTwoSMC)_{km}z_{2k}(\xTwoNew)
$$
and $(\xOneNew,\xTwoNew)$ is the new predictor pair that partners $\ynew$.

\section{Gaussian Response Models}\label{sec:GaussResp}

For reasons explained in Section \ref{sec:intro},  it is prudent to
first describe the online semiparametric regression via sequential Monte Carlo
for situations where Gaussianity of the responses can be assumed. We start with 
the familiar multiple linear regression setting.

\subsection{Multiple Linear Regression}

Let $\bX$ be a $n\times p$ design matrix and 
consider the Bayesian regression model
\begin{equation}
\by|\,\bbeta,\sigma^2\sim N(\bX\bbeta,\sigma^2\,\bI),\quad
\bbeta\sim N(\bmu_{\bbeta},\bSigma_{\bbeta}),\quad
\sigma\sim\mbox{Half-Cauchy}(\ssigsq)
\label{eq:linRegMod}
\end{equation}
As explained in Section \ref{sec:DistDefns} an equivalent, but more tractable model, 
is that where
$$\sigma\sim\mbox{Half-Cauchy}(\ssigsq)$$
is replaced by the auxiliary variable 
representation
\begin{equation}
\sigma^2|\,a\sim\mbox{Inverse-Gamma}(\smhalf,1/a),\quad
a\sim\mbox{Inverse-Gamma}(\smhalf,1/\ssigsq^2).
\label{eq:HCtoIG}
\end{equation}

Batch fitting of (\ref{eq:linRegMod}) via Markov chain Monte Carlo is very established
(e.g. Gelman \textit{et al.}, 2014; Chapters 11--12) and it is listed in Algorithm \ref{alg:batchMCMCforLM}.
It relies on the result (e.g. Tierney, 1994) that, after convergence can be assumed
following the ``burn-in'' phase of length $\Nburn$, 
\begin{equation}
\begin{array}{c}
\mbox{successive draws from the full conditional distributions of $\bbeta$, $\sigma^2$ and $a$}\\[0.2ex]
\mbox{constitute draws for the joint posterior density function: $\pDens(\bbeta,\sigma^2,a|\by)$.}
\end{array}
\label{eq:mainMCMCresult}
\end{equation}

%
%
\begin{algorithm}
\begin{center}
\begin{minipage}[t]{160mm}
\begin{itemize}
\item[] Data Inputs: $\by\ (n\times 1)$ and $\bX\ (n\times p)$. 
\item[] Markov Chain Monte Carlo Dimension Inputs: 
$\Nburn$ and $\Nkept$, both positive integers.\\[-3ex]
\item[] Hyperparameter Inputs: $\bmu_{\bbeta}\ (p\times 1)$, $\bSigma_{\bbeta}\ (p\times p)$
symmetric and positive definite,\ $\ssigsq>0$.
\item[] $\yTy\thickarrow\by^T\by$\ \ ;\ \ $\XTX\thickarrow\bX^T\bX$\ \ ;\ \ $\XTy\thickarrow\bX^T\by$
\item[] For $g=1,\ldots,\Nburn+\Nkept$:\\[-3ex]
\begin{itemize}
\item[] $\bOmega\thickarrow {\displaystyle\frac{\XTX}{(\sigma^2)^{[g-1]}}}+\bSigma_{\bbeta}^{-1}$
\ \ ;\ \ decompose $\bOmega=\bUOmega\diag(\bdOmega)\bUOmega^T$ where $\bUOmega\bUOmega^T=\bI$\\[-1ex]
\item[] $\bz\thickarrow \mbox{$p\times1$ vector containing totally independent $N(0,1)$ draws}$
\item[]$\bbeta^{[g]}\thickarrow 
\bUOmega\displaystyle{\left[\frac{\bUOmega^T\bz}{\sqrt{\bdOmega}}
+\frac{\bUOmega^T\Big\{\XTy\big/(\sigma^2)^{[g-1]}+\bSigma_{\bbeta}^{-1}\bmu_{\bbeta}\Big\}}
{\bdOmega}\right]}$\\[-1ex]
\item[] $a^{[g]}\simind\mbox{Inverse-Gamma}\Big(1,\{(\sigma^2)^{[g-1]}\}^{-1}
+\ssigsq^{-1}\Big)$ 
\item[]$(\sigma^2)^{[g]}\simind\mbox{Inverse-Gamma}\Big(\smhalf(n+1),(a^{[g]})^{-1}$\\
\null$\qquad\qquad\qquad\qquad+\smhalf\Big\{\yTy-2(\XTy)^T\bbeta^{[g]}
+\big(\bbeta^{[g]}\big)^T\XTX\bbeta^{[g]}\Big\}\Big)$.
\end{itemize}
\item[]
Produce summaries based on the kept $\bbeta^{[g]}$ and $(\sigma^2)^{[g]}$ chains,\\
$\Nburn+1\le g\le(\Nburn+\Nkept)$, being draws from the posterior distributions
of $\bbeta$ and $\sigma^2$ (due to result (\ref{eq:mainMCMCresult})).
\end{itemize}
\end{minipage}
\end{center}
\caption{\it Batch Markov chain Monte Carlo algorithm for approximate inference 
in the Gaussian response linear model.}
\label{alg:batchMCMCforLM}
\end{algorithm}
%
%

Note that Algorithm \ref{alg:batchMCMCforLM} uses the spectral decomposition of the 
matrix denoted by $\bOmega$ to efficiently obtain draws from the full conditional
distribution of the $\bbeta$ vector. The justifications for this and other aspects of 
Algorithm \ref{alg:batchMCMCforLM} are given in Section \ref{sec:batchLMjustif}
of the supplement.

Algorithm \ref{alg:onlineSMCforLM} is the online counterpart of 
Algorithm \ref{alg:batchMCMCforLM}, based on the general approach 
of Figure \ref{fig:genericFormSMC}. Algorithm \ref{alg:onlineSMCforLM} 
differs in that the data arrive sequentially and the fits and inferential summaries
are updated in real time. It has the attractive feature that
the posterior distribution updates depend only on
the sufficient statistics $\yTy$, $\XTy$ and $\XTX$. This implies 
that the streaming data does not have to be stored or used again
after the sufficient statistics have been updated.
In this sense, Algorithm \ref{alg:onlineSMCforLM} achieves purely online
fitting and inference according to the definition described in Section \ref{sec:intro}. 
Section \ref{sec:onlineSMCforLMjustif} of the supplement 
provides justifications for the Algorithm \ref{alg:onlineSMCforLM} steps.

%
%
\begin{algorithm}
\begin{center}
\begin{minipage}[h]{160mm}
\begin{itemize}
\item[] Tuning Parameter Inputs (defaults):\ $\npartic\in\naturalNumbers\ (1000)$\ ;\ 
$\tau>0\ (2/\npartic)$.
\item[] Hyperparameter Inputs: $\bmu_{\bbeta}\ (p\times 1)$, $\bSigma_{\bbeta}\ (p\times p)$
symmetric and positive definite,\ $\ssigsq>0$.
\item[] Initialize:
\begin{itemize}
\item[]$\bbetaSMC\thickarrow\mbox{$p\times\npartic$ matrix with columns containing independent random}$\\ 
$\null\ \ \qquad\qquad\mbox{samples from}\ N(\bmu_{\bbeta},\bSigma_{\bbeta})$
\item[]$\aSMC\thickarrow\mbox{$1\times\npartic$ vector containing a
random sample from Inverse-Gamma}(\smhalf,1/\seps^2)$%
\item[]$\sigsqSMC\thickarrow\mbox{$1\times\npartic$ vector containing a
random sample from Half-Cauchy}(\ssigsq)$
\item[]$\logw\thickarrow \log\big(\frac{1}{\npartic}\big)\bone$\ ;\ $n\thickarrow 0$
\ ;\ $\yTy\thickarrow0$
\ ;\ $\XTy\thickarrow\bzero\ (p\times1)$
\ ;\ $\XTX\thickarrow\bzero\ (p\times p)$.
\end{itemize}
\item[] Cycle:
\begin{itemize}
\item[] Read in $\ynew (1\times 1)$  and $\bxnew (p\times 1)$\ \ ;\ \ $n\thickarrow n+1$
\item[] $\yTy\thickarrow \yTy + \ynew^2$\ \ ;\ \
$\XTy\thickarrow \XTy + \bxnew\ynew$\ \ ;\ \ $\XTX\thickarrow \XTX + \bxnew\bxnew^T$
\item[] $\bdeta\thickarrow \bbetaSMC^T\bxnew$
\ \ \ ;\ \ \ $\logw\thickarrow\logw + (\ynew\bdeta - \smhalf\,\bdeta\odot\bdeta)/\{(\sigsqSMC)^T\}
-\smhalf\log\{(\sigsqSMC)^T\}$
\item[] $\bp\thickarrow\exp\{\logw - \max(\logw)\}/[\bone^T\exp\{\logw - \max(\logw)\}]$
\item[] If $\bp^T\bp>\tau$ then
\begin{itemize}
\item[] $\LMparmsSMC\thickarrow\SystematicResample\left(\LMparmsSMC,\bp\right)$
\ \ \ ;\ \ \   
$\logw\thickarrow\log\big(\frac{1}{\npartic}\big)\bone$
\end{itemize}
\item[] For $m=1,\ldots,\npartic$:
\begin{itemize}
\item[] $\bOmega\thickarrow {\displaystyle\frac{\XTX}{(\sigsqSMC)_m}}+\bSigma_{\bbeta}^{-1}$
\ \ ;\ \ decompose $\bOmega=\bUOmega\diag(\bdOmega)\bUOmega^T$ where $\bUOmega\bUOmega^T=\bI$\\[-1ex]
\item[] $\bz\thickarrow \mbox{$p\times 1$ vector containing totally independent $N(0,1)$ draws}$
\item[] $m$th column of $\bbetaSMC\thickarrow\bUOmega\displaystyle{\left\{\frac{\bUOmega^T\bz}{\sqrt{\bdOmega}}
+\frac{\bUOmega^T\Big(\XTy\big/(\sigsqSMC)_m+\bSigma_{\bbeta}^{-1}\bmu_{\bbeta}\Big)}
{\bdOmega}\right\}}$
\item[] $(\aSMC)_m\sim\mbox{Inverse-Gamma}\Big(1,(\sigsqSMC)_m^{-1}+\ssigsq^{-1}\Big)$
\item[]$(\sigsqSMC)_m\sim\mbox{Inverse-Gamma}\Big(\smhalf(n+1),(\aSMC)_m^{-1}$\\
\null$\qquad\qquad\qquad\qquad+\smhalf\Big\{\yTy-2\big((\XTy)^T\bbetaSMC\big)_m
+\big(\bbetaSMC^T\XTX\bbetaSMC\big)_{mm}
\Big\}\Big).$
\end{itemize}
\item[] Produce summaries based on the current approximate posterior distributions of $\bbeta$ 
and $\sigma^2$ equalling the probability mass functions with atoms stored in $\bbetaSMC$ and $\sigsqSMC$, 
respectively, and probabilities $\bp$.
\end{itemize}
\item[] until data no longer available or analysis terminated.
\end{itemize}
\end{minipage}
\end{center}
\caption{\it Online sequential Monte Carlo algorithm for online approximate inference 
in the Gaussian response linear model.}
\label{alg:onlineSMCforLM} 
\end{algorithm}
%
%

Algorithm \ref{alg:onlineSMCforLMwTune} is a modification
of Algorithm \ref{alg:onlineSMCforLM} that allows for the
possibility of batched-based tuning at the start of the 
online regression analysis. Its justification is given 
in Section \ref{sec:onlineSMCforLMwTuneJustif} of the supplement.
An illustration of batch-based tuning to properly initialise 
an online semiparametric regression analysis is given
in Section \ref{sec:binaryNPRexamp} (see Figure \ref{fig:nWarmBinNPR}).

%
%
\begin{algorithm}
\begin{center}
\begin{minipage}[t]{160mm}
\begin{itemize}
\item[1.] Set $\nwarm$ to be the warm-up sample size
and $\nvalid$ to be size of the validation period.
Read in the first $\nwarm+\nvalid$ response and predictor
values.
\item[2.] Create $\bywarm$ and $\bXwarm$ consisting of the
first $\nwarm$ response and predictor values.
\item[3.]Feed $\bywarm$ and $\bXwarm$ into the batch Markov chain Monte Carlo Algorithm 
\ref{alg:batchMCMCforLM} with $\Nkept=M$. Use the kept chains $\bbeta^{[g]}$, 
$(\sigma^2)^{[g]}$ and $a^{[g]}$, $1\le g\le M$,
to initialise $\bbetaSMC$, $\aSMC$ and $\sigsqSMC$.
\item[4.] Set $\yTy\leftarrow \bywarm^T\bywarm$,
\ $\XTy\leftarrow \bXwarm^T\bywarm$,
\ $\XTX\leftarrow \bXwarm^T\bXwarm$ and $n\leftarrow \nwarm$.
\item[5.] Run the online sequential Monte Carlo Algorithm \ref{alg:onlineSMCforLM}
until $n=\nwarm+\nvalid$.
\item[6.] Use convergence diagnostic graphics to assess whether
the online parameters are converging to the batch parameters.
\begin{itemize}
\item[(a)] If not converging then return to Step 1 and 
increase $\nwarm$. 
\item[(b)] If converging then continue running the online sequential Monte Carlo Algorithm 
\ref{alg:onlineSMCforLM} until data no longer available or analysis terminated.
\end{itemize}
\end{itemize}
\end{minipage}
\end{center}
\caption{\it Modification of Algorithm \ref{alg:onlineSMCforLM}
to include batch-based tuning and convergence diagnosis.}
\label{alg:onlineSMCforLMwTune} 
\end{algorithm}
%
%

Algorithm \ref{alg:onlineSMCforLM} can be used to produce
convergence diagnostic graphics analogous to 
Figures 3 and 5 in Luts, Broderick \myand Wand (2014).

\subsection{Linear Mixed Models}

Our mixed model-based splines approach to Bayesian semiparametric regression 
makes use of the following class of linear mixed models: 
\begin{equation}
\begin{array}{c}
\by|\,\bbeta,\bu,\sigeps^2 \sim 
N(\bX\bbeta+\bZ\bu,\sigeps^2\,\bI),\\[3ex]
\bu|\,\sigma_{u1}^2,
\ldots,\sigma_{uR}^2\sim N(\bzero,\mbox{blockdiag}
(\sigma_{u1}^2\,\bI_{K_1},\ldots,\sigma_{uR}^2\,\bI_{K_R})).
\end{array}
\label{eq:LMM1}
\end{equation}
Here $\by$ is an $n\times1$ vector of response variables, 
$\bbeta$ is a $p\times1$ vector of fixed effects, 
$\bu$ is a vector of random effects, $\bX$ and $\bZ$ 
are design matrices, $\sigeps^2$ is the error
variance and $\sigma_{u1}^2,\ldots,\sigma_{ur}^2$
are variance parameters corresponding to sub-blocks
of $\bu$ of size $K_1,\ldots,K_R$.
We set the priors to be
\begin{equation}
\bbeta\sim N(\bmu_{\bbeta},\bSigma_{\bbeta}),\quad
\sigma_{ur}\sim\mbox{Half-Cauchy}(\sur),\ 1\le r\le R,\quad 
\sigeps\sim\mbox{Half-Cauchy}(\seps)
\label{eq:LMM2}
\end{equation}
with the hyperparameters $\bSigma_{\bbeta}$ symmetric and positive definite and 
$\seps,\sur>0$ for $1\le r\le R$. As in  Section \ref{sec:GaussResp}, 
we introduce the auxiliary variables 
\begin{equation}
a_{ur}\sim\mbox{Inverse-Gamma}(\smhalf,1/\sur^2)
\quad\mbox{and}\quad
a_{\varepsilon}\sim\mbox{Inverse-Gamma}(\smhalf,1/\seps^2)
\label{eq:LMM3}
\end{equation}
and use of the analogue of (\ref{eq:HCtoIG}) to induce Half-Cauchy
priors on the standard deviation parameters.

As described in Section 2 of Zhao, Staudenmayer, Coull \myand Wand (2006),
model (\ref{eq:LMM1})--(\ref{eq:LMM2}) covers several
important special cases, including (with example number from 
Zhao {\it et al.}\ 2006 added):
\begin{itemize}
\item simple random effects models (Examples 1 and 2),
\item cross random effects models (Example 3),
\item nested random effects models (Example 4),
\item generalized additive models (Example 6),
\item semiparametric mixed models (Example 7),
\item bivariate smoothing and geoadditive models 
extensions (Example 8).
\end{itemize}
Examples 2 and 6 of Zhao {\it et al.}\ (2006) involve $2\times2$ and
$3\times3$ unstructured covariance matrix parameters which, strictly 
speaking, are not special cases of (\ref{eq:LMM2}).
However, as discussed in Section \ref{sec:unstrucCov}, 
the unstructured covariance matrix extension is quite straightforward.

Let
$$\bC=[\bX\ \bZ]$$
be the combined design matrix in (\ref{eq:LMM1}) and $P$ be the number of columns in $\bC$. 
Then each pass of the corresponding online sequential Monte Carlo algorithm involves 
arrival and processing of a new scalar response measurement, $\ynew$, and a
$P\times 1$ vector $\bcnew$, corresponding to the new row of $\bC$.
This results in Algorithm \ref{alg:onlineSMCforLMM} for purely online fitting of (\ref{eq:LMM1}). 
Its justification is given in Section \ref{sec:onlineSMCforLMMjustif} of the supplement.

%
%
\begin{algorithm}
\begin{center}
\begin{minipage}[h]{151mm}
\begin{itemize}
\item[] \item[] Tuning Parameter Inputs (defaults):\ 
$\npartic\in\naturalNumbers\ (1000)$\ ;\ $\tau>0\ (2/\npartic)$.
\item[] Hyperparameter Inputs: $\bmu_{\bbeta}\ (p\times 1)$, $\bSigma_{\bbeta}\ (p\times p)$
symmetric and positive definite,\ $\ssigsq>0$, $\sur>0,\ 1\le r\le R$.
\item[] Perform batch-based tuning runs analogous to those described in 
Algorithm \ref{alg:onlineSMCforLMwTune} and determine a warm-up sample size $\nwarm$ for 
which convergence is validated. 
\item[] Set $\bywarm$ and $\bCwarm$ to be the response vector
and design matrix based on the first $\nwarm$ observations.
Then set
\ $\yTy\leftarrow \bywarm^T\bywarm$,\ 
\ $\CTy\leftarrow \bCwarm^T\bywarm$,
\ $\CTC\leftarrow \bCwarm^T\bCwarm$,
\ $n\leftarrow\nwarm$. 
\item[] Set the following matrices (with dimensions)
$$\bbetabuSMC\ (P\times M)\ ;\ \sigsqEpsSMC\ (1\times M)\ ;\ \aEpsSMC\ (1\times M)
\ ;\ \bsigsqUSMC\ (r\times M)\ ; \ \baUSMC\ (r\times M)$$
such that each column is a random sample from the relevant approximate posterior distribution
according to the batch Markov chain Monte Carlo samples based on the first $\nwarm$ observations.
\item[] Cycle:
\begin{itemize}
\item[] Read in $\ynew\ (1\times 1)$ and $\bcnew\ (P\times 1)$
\ \ ;\ \ $n\leftarrow n+1$
\item[] $\bdeta\thickarrow\bbetabuSMC^T\bcnew$
\ \ \ ;\ \ \ $\logw\thickarrow\logw + (\ynew\bdeta - \smhalf\,\bdeta\odot\bdeta)/\{(\sigsqEpsSMC)^T\}
-\smhalf\log\{(\sigsqEpsSMC)^T\}$
\item[] $\bp\thickarrow\exp\{\logw - \max(\logw)\}/[\bone^T\exp\{\logw - \max(\logw)\}]$
\item[] If $\bp^T\bp>\tau$ then
\begin{itemize}
\item[] $\LMMparmsSMC\thickarrow\SystematicResample\left(\LMMparmsSMC,\bp\right)$
\ ;\ $\logw\thickarrow\log\big(\frac{1}{\npartic}\big)\bone$
\end{itemize}
\item[] $\yTy\leftarrow\yTy+\ynew^2$\ \ ;\ \ 
$\CTy\leftarrow\CTy+\bcnew\,\ynew$\ \ ;\ \
$\CTC\leftarrow\CTC+\bcnew\,\bcnew^T$
\item[] For $m=1,\ldots,\npartic$:
\begin{itemize}
\item[] $\bOmega\thickarrow {\displaystyle\frac{\CTC}{(\sigsqEpsSMC)_m}}
+\mbox{blockdiag}\Big(\bSigma_{\bbeta}^{-1},
\bI_{K_1}/(\bsigsqUSMC)_{1m},\ldots,\bI_{K_R}/(\bsigsqUSMC)_{Rm}\Big)$\\[1ex]
decompose $\bOmega=\bUOmega\diag(\bdOmega)\bUOmega^T$ where $\bUOmega\bUOmega^T=\bI$\\[-1ex]
\item[] $\bz\thickarrow \mbox{$P\times1$ vector containing totally independent $N(0,1)$ draws}$
\item[] $m$th column of $\bbetabuSMC\thickarrow 
\bUOmega\displaystyle{\left[\frac{\bUOmega^T\bz}{\sqrt{\bdOmega}}
+\frac{\bUOmega^T\Big\{\CTy\big/(\sigsqEpsSMC)_m+\bSigma_{\bbeta}^{-1}\bmu_{\bbeta}\Big\}}
{\bdOmega}\right]}$\\[-1ex]
\item[] \textit{continued on a subsequent page $\ldots$}\\[-2ex]
\end{itemize}
\end{itemize}
\end{itemize}
\end{minipage}
\end{center}
\caption{\it Online sequential Monte Carlo algorithm for approximate inference 
in the Gaussian response linear mixed model (\ref{eq:LMM1}).}
\label{alg:onlineSMCforLMM}
\end{algorithm}
%
%

\setcounter{algorithm}{4}
%
%
\begin{algorithm}
\begin{center}
\begin{minipage}[h]{151mm}
\begin{itemize}
\item[] 
\begin{itemize}
\item[] 
\begin{itemize}
\item[] $(\aEpsSMC)_m\sim\mbox{Inverse-Gamma}\left(1,(\sigsqEpsSMC)_m^{-1}
+\seps^{-1}\right)$
\item[]$(\sigsqEpsSMC)_m\sim\mbox{Inverse-Gamma}\Bigg(\smhalf(n+1),(\aEpsSMC)_m^{-1}$\\
\null$\qquad+\smhalf\left\{\yTy-2\left((\CTy)^T\bbetabuSMC\right)_m
+\left(\bbetabuSMC^T\CTC\bbetabuSMC\right)_{mm}
\right\}\Bigg)$
\item[] $\iStt\thickarrow1$
\item[] For $r=1,\ldots,R$\,:
\begin{itemize}
\item[] $(\baUSMC)_{rm}\sim\mbox{Inverse-Gamma}\left(1,(\bsigsqUSMC)_{rm}^{-1}
+\sur^{-1}\right)$
\item[] $\iEnd\thickarrow\iStt+K_r-1$\ ;\  
$\bomega\thickarrow \mbox{entries $\iStt$ to $\iEnd$ of the $m$th column of $\buSMC$}$
\item[]$\iStt\thickarrow\iEnd+1$
\item[] $(\bsigsqUSMC)_{rm}\sim\mbox{Inverse-Gamma}\left(\smhalf(K_r+1),
(\baUSMC)_{rm}^{-1}+\smhalf\Vert\bomega\Vert^2\right)$.
\end{itemize}
\end{itemize}
\item[] Produce summaries based on the current approximate posterior distributions of 
$\bbetabu$, $\sigeps^2$ and $(\sigma^2_{u1},\ldots,\sigma^2_{uR})$ equalling the 
probability mass functions with atoms stored in $\bbetabuSMC$, $\sigsqEpsSMC$ 
$\bsigsqUSMC$ respectively, and probabilities $\bp$.
\end{itemize}
\item[] until data no longer available or analysis terminated.
\end{itemize}
\end{minipage}
\end{center}
\caption{\textbf{continued.}\ \textit{This is a continuation of the description of this algorithm that
continues on a preceding page.}}
\end{algorithm}
%
%

\subsection{Extension to Unstructured Covariance Matrices for Random Effects}
\label{sec:unstrucCov}

A simple special case of (\ref{eq:LMM1}) is the \emph{random intercept model}, 
for which the first two hierarchical levels are set to
\begin{equation}
\begin{array}{c}
y_{ij}|\beta_0,\beta_1,U_i,\sigeps^2\simind
N(\beta_0+U_i+\beta_1\,x_{ij},\sigeps^2),\quad 1\le i\le m,
\quad 1\le j\le n_i,\\[2ex]
\mbox{and}\quad U_i\big|\sigma_u^2\simind N(0,\sigma_u^2).
\end{array}
\label{eq:randInt}
\end{equation}
The \emph{random intercepts and slopes} extension of (\ref{eq:randInt}) is
$$
\begin{array}{c}
y_{ij}|\beta_0,\beta_1,U_i,V_i,\sigeps^2\simind
N(\beta_0+U_i+(\beta_1+V_i)\,x_{ij},\sigeps^2),\quad 1\le i\le m,
\quad 1\le j\le n_i,\\[2ex]
\mbox{and}\quad
\left[
\begin{array}{c}
U_i\\
V_i \\
\end{array}
\right]\Big|\bSigma\simind N(\bzero,\bSigma),
\quad\mbox{where}\quad
\bSigma\equiv  \left[\begin{array}{cc}         
\sigma_u^2 & \rho_{uv}\,\sigma_u\,\sigma_v \\
\rho_{uv}\,\sigma_u\,\sigma_v & \sigma_v^2
\end{array}\right]
\end{array}
$$
is an unstructured $2\times2$ covariance matrix.
The conjugate prior for $\bSigma$ is the Inverse Wishart distribution.
Note that
$$
\begin{array}{c}
\bSigma|\,a_{uv1},a_{uv2}\sim\mbox{Inverse-Wishart}\left(\nu+1,2\nu\,
\left[
\begin{array}{cc}
1/a_{uv1} & 0 \\
   0  & 1/a_{uv2} 
\end{array}
\right]
\right),\\[2ex]
a_{uv1},a_{uv2}\simind
\mbox{Inverse-Gamma}(\smhalf,1/s_{uv}),\quad \nu,s_{uv}>0
\end{array}
$$
provides a covariance matrix extension
of $\sigma_u\sim\mbox{Half-Cauchy}(A_u)$.
The choice $\nu=2$ imposes a $\mbox{Uniform}(-1,1)$ distribution on $\rho_{uv}$
and Half-$t_2$ distributions on $\sigma_u$ and $\sigma_v$.
This is described in Huang \myand Wand (2013), including
the definition of the $\mbox{Inverse-Wishart}(a,\bB)$ 
distribution. Extensions to models with larger unstructured covariance 
matrices is similar.

\section{Generalized Response Models}\label{sec:generResp}

We now switch attention to semiparametric regression models for which the
response is non-Gaussian, which we will refer to as \emph{generalized} response
models, and include binary and count response types. We begin with 
generalized linear models, for which the gist of the generalized response
extension can be conveyed without a big notational burden.

\subsection{Generalized Linear Models}

As with the (\ref{eq:linRegMod}) set-up, let $\bX$ be a $n\times p$ design matrix 
and $\bbeta$ be a $p\times 1$ coefficient vector. In this
subsection we now suppose that the entries of the $n\times1$ response vector $\by$ 
have the following one-parameter exponential family probability mass or
density function:
\begin{equation}
\pDens(\by|\bbeta)
=\exp\big\{\by^T\bX\bbeta-\bone^Tb\big(\bX\bbeta\big)+\bone^Tc(\by)\big\}h(\by)
\label{eq:GLMbasic}
\end{equation}
for particular scalar-to-scalar functions $b$, $c$ and $h$ with the convention
that function evaluation is applied in an element-wise fashion.
The logistic regression special case, for binary response data, corresponds to 
\begin{equation}
b(x)=\log(e^x+1),\quad c(x)=0\quad\mbox{and}\quad h(x)=I(x\in\{0,1\}).
\label{eq:bchLogis}
\end{equation}
Instead, setting
\begin{equation}
b(x)=e^x,\quad c(x)=-\log(x!)\quad\mbox{and}\quad h(x)=I(x\in\{0,1,2,\ldots\})
\label{eq:bchPoiss}
\end{equation}
corresponds to Poisson regression for count responses.

Online fitting of (\ref{eq:GLMbasic}) is provided by Algorithm \ref{alg:onlineSMCforGLM}
and justified in Section \ref{sec:NeddieSeagoon} of the supplement.
An important difference between Algorithm \ref{alg:onlineSMCforGLM}
for generalized linear models and  Algorithm \ref{alg:onlineSMCforLM} for Gaussian
response linear models is that purely online fitting is \emph{not} being achieved.
Recall that Algorithm \ref{alg:onlineSMCforLM} is such that the data can 
be discarded after each sufficient statistic update is accomplished. 
In contrast, Algorithm \ref{alg:onlineSMCforGLM} is such that,
every time a new data vector arrives the full data to date needs to be available 
and processed for the approximate posterior distribution updates.
The essence of this difference is the non-Gibbsian nature of the $\bbeta$ vector
full conditional distribution in the generalized response situation. Instead
of the simple closed form update that arises in the Gaussian case, a Metropolis-Hastings
scheme has to be called up. The logarithm of Metropolis-Hastings ratio, denoted
in Algorithm \ref{alg:onlineSMCforGLM} by $\lambda$, requires the full data to 
date.

\subsubsection{Choice of the Metropolis-Hastings Random Walk Scale Parameter}\label{sec:upsilonChoice}

Algorithm \ref{alg:onlineSMCforGLM} involves the following step:
\begin{equation}
\bbetaRW\thickarrow \bbetaSMCm+\displaystyle{\frac{\upsilon\bz}{\sqrt{n}}}
\label{eq:SqueakyFromme}
\end{equation}
for some choice of the scale parameter $\upsilon>0$. As explained in 
Section \ref{sec:NeddieSeagoon} of the supplement, (\ref{eq:SqueakyFromme}) corresponds to 
drawing from random walk proposal distribution as part of a Metropolis-Hastings
scheme for obtaining a draw from the current full conditional distribution of $\bbeta$.

Several strategies to select $\upsilon$ have been developed.
Sophisticated approaches such as the Metro\-polis-adjusted Langevin algorithm
(Roberts \myand Stramer, 2003) introduce also a deterministic drift and exploit gradient information
to provide principled choices of $\upsilon$, which can be further refined by considering higher-order
derivatives (Girolami \myand Calderhead, 2011).
Fearnhead and Taylor (2013) were among the
first to bring ideas from adaptive Markov chain Monte Carlo to bear in the context of sequential
Monte Carlo, and suggested to adapt $\upsilon$ based on the expected squared jumping distance; 
see also Bon \textit{et al.} (2021).
Maximising the expected squared jumping distance is equivalent to minimising the first-order autocorrelation of
the Markov chain and computation is straightforward.
Chopin and Papaspiliopoulos (2020, Section 17.2.1) recommend to use an estimate
of the sample covariance from the previous step of sequential Monte Carlo to calibrate the covariance
matrix of a general multivariate Gaussian proposal.  However, despite the potential
for these approaches to deliver improved mixing, they can also introduce novel
failure modes. For example, expected squared jumping distance may not be concave as $\upsilon$ is varied, 
which means that optimisation could, in principle, be difficult. To promote robustness, here
we use a simpler approach, based on theoretical results in  Roberts \myand Rosenthal (2001).
This entails choosing $\upsilon$ so that about 23\% of the particles are updated according to the Algorithm 
\ref{alg:onlineSMCforGLM} step:
$$\mbox{if $\lambda>\log(u)$ then the $m$th column of $\bbetaSMC$}\thickarrow\bbetaRW\ (m=1,\ldots,M).$$
The ``about 23\% of particles updated'' approach to setting $\upsilon$ values can be
done using the warm-up phase, and also involve simple-to-implement
adaptations to $\upsilon$ as the data stream in.  
The illustrations in Sections \ref{sec:logisRegnExamp} and \ref{sec:binaryNPRexamp} 
use such an approach for choice of $\upsilon$.

%
%
\begin{algorithm}
\begin{center}
\begin{minipage}[t]{160mm}
\begin{itemize}
\item[] Tuning Parameter Inputs (defaults):\ $\npartic\in\naturalNumbers\ (1000)$
\ ;\ $\tau>0\ (2/\npartic)$
\ ; \ $\upsilon>0$ (see Section \ref{sec:upsilonChoice}).
\item[] Hyperparameter Inputs: $\bmu_{\bbeta}\ (p\times 1)$, $\bSigma_{\bbeta}\ (p\times p)$
symmetric and positive definite.
\item[] Perform batch-based tuning runs analogous to those described in 
Algorithm \ref{alg:onlineSMCforLMwTune} and determine a warm-up sample size $\nwarm$ for 
which convergence is validated. 
\item[] Set $\bywarm$ and $\bXwarm$ to be the response vector
and design matrix based on the first $\nwarm$ observations.
Then set $n\thickarrow\nwarm$, $\by\thickarrow\bywarm$ and $\bX\thickarrow\bXwarm$.
\item[] Cycle:
\begin{itemize}
\item[] Read in $\ynew (1\times 1)$  and $\bxnew (p\times 1)$\ \ ;\ \ $n\thickarrow n+1$
\item[] $\bdeta\thickarrow \bbetaSMC^T\bxnew$
\ \ \ ;\ \ \ $\logw\thickarrow\logw + \ynew\bdeta - b(\bdeta)$
\item[] $\bp\thickarrow\exp\{\logw - \max(\logw)\}/[\bone^T\exp\{\logw - \max(\logw)\}]$
\item[] If $\bp^T\bp>\tau$ then
\begin{itemize}
\item[] $\bbetaSMC\thickarrow\SystematicResample\big(\bbetaSMC,\bp\big)$
\ \ \ ;\ \ \   
$\logw\thickarrow\log\big(\frac{1}{\npartic}\big)\bone$\\
\end{itemize}
\item[] $\by\thickarrow\left[\begin{array}{c}   
\by\\[1ex]
\ynew
\end{array}
\right]$\ \ ;\ \ $\bX\thickarrow\left[\begin{array}{c}   
\bX\\[1ex]
\bxnew^T
\end{array}
\right]$
\item[] For $m=1,\ldots,\npartic$:
\begin{itemize}
\item[] $\bz\thickarrow p\times1\ \mbox{vector containing totally independent $N(0,1)$ draws}$
\item[] $\bbetaSMCm\thickarrow\mbox{$m$th column of $\bbetaSMC$}$\ \ ;\ \ 
$\bbetaRW\thickarrow \bbetaSMCm+\displaystyle{\frac{\upsilon\bz}{\sqrt{n}}}$
\item[] $\bdetaSMC\thickarrow \bX\bbetaSMCm$\ \ ;\ \ $\bdetaRW\thickarrow\bX\bbetaRW$
\item[] $\lambda\thickarrow \by^T(\bdetaRW-\bdetaSMC)-\bone^T\{b(\bdetaRW)-b(\bdetaSMC)\}$\\
        \null$\qquad\qquad-\smhalf\bbetaRW^T\bSigma_{\bbeta}^{-1}\bbetaRW
          +\smhalf\bbetaSMCm^T\bSigma_{\bbeta}^{-1}\bbetaSMCm
          +(\bbetaRW-\bbetaSMCm)^T\bSigma_{\bbeta}^{-1}\bmu_{\bbeta}$
\item[] $u\thickarrow \mbox{draw from the Uniform$(0,1)$ distribution}$
\item[] if $\lambda>\log(u)$ then
\begin{itemize}
\item[] $\mbox{$m$th column of $\bbetaSMC$}\thickarrow\bbetaRW$
\end{itemize}
\end{itemize}
\item[] Produce summaries based on the current approximate posterior distribution of $\bbeta$ 
equalling the probability mass functions with atoms stored in $\bbetaSMC$ and probabilities $\bp$.
\end{itemize}
\item[] until data no longer available or analysis terminated.
\end{itemize}
\end{minipage}
\end{center}
\caption{\it Online sequential Monte Carlo algorithm for online approximate inference 
in the generalized response linear model.}
\label{alg:onlineSMCforGLM} 
\end{algorithm}
%
%

\subsection{Generalized Linear Mixed Models}

The class of Bayesian generalized linear mixed models which we consider is
\begin{equation}
\begin{array}{c}
\pDens(\by|\,\bbeta,\bu)=
\exp\big\{\by^T(\bX\bbeta+\bZ\bu)-\bone^Tb(\bX\bbeta+\bZ\bu)+\bone^Tc(\by)\big\}h(\by)\\[1.5ex]
\bu|\,\sigma_{u1}^2,
\ldots,\sigma_{uR}^2\sim N(\bzero,\mbox{blockdiag}
(\sigma_{u1}^2\,\bI_{K_1},\ldots,\sigma_{uR}^2\,\bI_{K_r}))
\end{array}
\label{eq:GLMM1}
\end{equation}
where $\bbeta$ and $\sigma_{ur}^2$, $1\le r\le R$, have prior distributions
as given by (\ref{eq:LMM2}). Model (\ref{eq:GLMM1}) has similar utility to (\ref{eq:LMM1}) 
for various semiparametric regression scenarios, but for generalized response situations. In particular, 
logistic mixed models and Poisson mixed models correspond to the $b$, $c$ and $h$ functions 
given by (\ref{eq:bchLogis}) and (\ref{eq:bchPoiss}), respectively.

Algorithm \ref{alg:onlineSMCforGLMM} describes online fitting of (\ref{eq:GLMM1}) 
via sequential Monte Carlo, with justification provided by Section 
\ref{sec:onlineSMCforGLMMjustif} of the supplement. 
An illustration of Algorithm \ref{alg:onlineSMCforGLMM} is given
in Section \ref{sec:binaryNPRexamp}.

%
%
\begin{algorithm}
\begin{center}
\begin{minipage}[h]{151mm}
\begin{itemize}
\item[] Tuning Parameter Inputs (defaults):\ $\npartic\in\naturalNumbers\ (1000)$
\ ;\ $\tau>0\ (2/\npartic)$\ ; \ $\upsilon$ (see Section \ref{sec:upsilonChoice}).
\item[] Hyperparameter Inputs: $\bmu_{\bbeta}\ (p\times 1)$, $\bSigma_{\bbeta}\ (p\times p)$
symmetric and positive definite, $\sur>0$, $1\le r\le R$. 
\item[] Perform batch-based tuning runs analogous to those described in 
Algorithm \ref{alg:onlineSMCforLMwTune} and determine a warm-up sample size $\nwarm$ for 
which convergence is validated. 
\item[] Set $\bywarm$ and $\bCwarm$ to be the response vector
and design matrix based on the first $\nwarm$ observations.
\item[] Set the following matrices (with dimensions)
$$\bbetabuSMC\ (P\times M)\quad;\quad\bsigsqUSMC\ (r\times M)\quad;\quad\baUSMC\ (r\times M)$$
such that each column is a random sample from the relevant approximate posterior distribution
according to the batch Markov chain Monte Carlo samples based on the first $\nwarm$ observations.
\item[] Cycle:
\begin{itemize}
\item[] Read in $\ynew\ (1\times 1)$ and $\bcnew\ (P\times 1)$
\ \ ;\ \ $n\leftarrow n+1$
\item[] $\bdeta\thickarrow\bbetabuSMC^T\bcnew$
\ \ \ ;\ \ \ $\logw\thickarrow\logw + \ynew\bdeta - b(\bdeta)$
\item[] $\bp\thickarrow\exp\{\logw - \max(\logw)\}/[\bone^T\exp\{\logw - \max(\logw)\}]$
\item[] If $\bp^T\bp>\tau$ then
\begin{itemize}
\item[] $\GLMMparmsSMC\thickarrow\SystematicResample\left(\GLMMparmsSMC,\bp\right)$
\ ;\ $\logw\thickarrow\log\big(\frac{1}{\npartic}\big)\bone$
\end{itemize}
\item[] $\by\thickarrow\left[\begin{array}{c}   
\by\\[1ex]
\ynew
\end{array}
\right]$\ \ ;\ \ $\bC\thickarrow\left[\begin{array}{c}   
\bC\\[1ex]
\bcnew^T
\end{array}
\right]$
\item[] For $m=1,\ldots,\npartic$:
\begin{itemize}
\item[] $\bz\thickarrow P\times1\ \mbox{vector containing totally independent $N(0,1)$ draws}$
\item[] $\bbetabuSMC_m\thickarrow\mbox{$m$th column of $\bbetabuSMC$}$\ \ ;\ \ 
$\bbetabuRW\thickarrow \bbetabuSMC_m+\displaystyle{\frac{\upsilon\bz}{\sqrt{n}}}$
\item[] $\bdetaSMC\thickarrow \bC\bbetabuSMC_m$\ \ ;\ \ $\bdetaRW\thickarrow\bC\bbetabuRW$
\item[] $\lambda\thickarrow \by^T(\bdetaRW-\bdetaSMC)-\bone^T\{b(\bdetaRW)-b(\bdetaSMC)\}$\\[1.5ex]
        \null$\qquad\qquad-\smhalf\bbetaRW^T\bSigma_{\bbeta}^{-1}\bbetaRW
          +\smhalf\bbetaSMCm^T\bSigma_{\bbeta}^{-1}\bbetaSMCm
          +(\bbetaRW-\bbetaSMCm)^T\bSigma_{\bbeta}^{-1}\bmu_{\bbeta}$
\item[] $\iStt\thickarrow1$
\item[] For $r=1,\ldots,R$:
\begin{itemize}
\item[] $\iEnd\thickarrow\iStt+K_r-1$\ ;
\item[] $\bomegaSMC\thickarrow \mbox{entries $\iStt$ to $\iEnd$ of the $m$th column of $\buSMC$}$
\item[]$\bomegaRW\thickarrow \mbox{entries $\iStt$ to $\iEnd$ of the $m$th column of $\buRW$}$
\end{itemize}
\item[] \textit{continued on a subsequent page $\ldots$}\\[-2ex]
\end{itemize}
\end{itemize}
\end{itemize}
\end{minipage}
\end{center}
\caption{\it Online sequential Monte Carlo algorithm for approximate inference 
in the generalized linear mixed model (\ref{eq:GLMM1}).}
\label{alg:onlineSMCforGLMM}
\end{algorithm}
%
%

\setcounter{algorithm}{6}
%
%
\begin{algorithm}
\begin{center}
\begin{minipage}[h]{151mm}
\begin{itemize}
\item[] $\iStt\thickarrow\iEnd+1$
\item[]$\lambda\thickarrow\lambda-
\smhalf\big(\Vert\bomegaSMC\Vert^2-\Vert\bomegaRW\Vert^2\big)\Big/(\bsigsqUSMC)_{rm}$
\item[] $u\thickarrow \mbox{draw from the Uniform$(0,1)$ distribution}$
\item[] if $\lambda>\log(u)$ then
\begin{itemize}
\item[] $\mbox{$m$th column of $\bbetabuSMC$}\thickarrow\bbetabuRW$
\begin{itemize}
\item[] $\iStt\thickarrow1$
\item[] For $r=1,\ldots,R$\,:
\begin{itemize}
\item[] $(\baUSMC)_{rm}\sim\mbox{Inverse-Gamma}\left(1,(\bsigsqUSMC)_{rm}^{-1}
+\sur^{-1}\right)$
\item[] $\iEnd\thickarrow\iStt+K_r-1$\ ;\  
$\bomega\thickarrow \mbox{entries $\iStt$ to $\iEnd$ of the $m$th column of $\buSMC$}$
\item[]$\iStt\thickarrow\iEnd+1$
\item[] $(\bsigsqUSMC)_{rm}\sim\mbox{Inverse-Gamma}\left(\smhalf(K_r+1),
(\baUSMC)_{rm}^{-1}+\smhalf\Vert\bomega\Vert^2\right)$.
\end{itemize}
\end{itemize}
\item[] Produce summaries based on the current approximate posterior distributions of 
$\bbetabu$ and $(\sigma^2_{u1},\ldots,\sigma^2_{uR})$ equalling the 
probability mass functions with atoms stored in $\bbetabuSMC$ and
$\bsigsqUSMC$ respectively, and probabilities $\bp$.
\end{itemize}
\item[] until data no longer available or analysis terminated.
\end{itemize}
\end{minipage}
\end{center}
\caption{\textbf{continued.}\ \textit{This is a continuation of the description of 
this algorithm that continues on a preceding page.}}
\end{algorithm}
%
%

\null\vfill\eject
\section{Illustrations}\label{sec:illus}

We have tested Algorithms \ref{alg:onlineSMCforLM}--\ref{alg:onlineSMCforGLMM} on 
many simulated and actual data sets. In this section we give some illustrations
of the practical performance of the new methodology. The first one
includes a comparison with the Luts \textit{et al.} (2014) variational approach.

\subsection{Online Logistic Regression}\label{sec:logisRegnExamp}

Algorithm 5 of Luts \textit{et al.} (2014) offers real-time logistic regression
with pure online updating based on the logistic log-likelihood variational
approximations of Jaakkola \myand Jordan (2000). However, as we mentioned in
Section \ref{sec:intro}, this online mean field variational Bayes approach 
to logistic regression is susceptible to poor accuracy. This problem, and
the online sequential Monte Carlo remedy, is illustrated here via a simple
linear logistic regression scenario.

\begin{figure}[!h]
\centering
\includegraphics[width=1.0\textwidth]{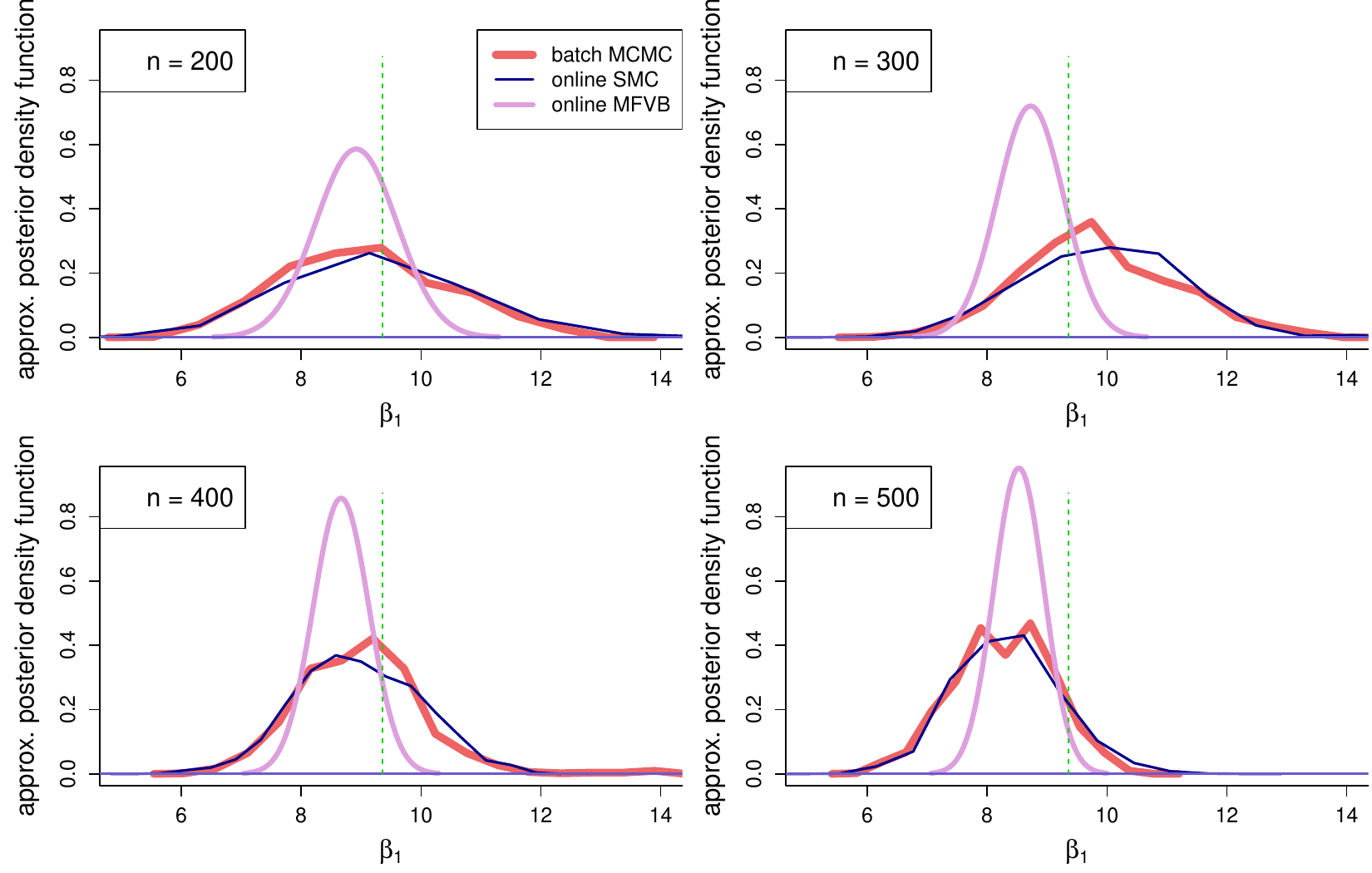}
\caption{\textit{Comparison of the approximations of $\pDens(\beta_1|\bxcurr,\bycurr)$
for the online logistic regression example.
The batch Markov chain Monte Carlo (MCMC) approximation is displayed as
a frequency polygon density estimate based on a kept MCMC sample of size
$1,000$ and bin width given by (\ref{eq:fpBWrule}). 
The online sequential Monte Carlo (SMC) approximation is 
displayed as a frequency polygon representation of a probability
mass function with $1,000$ atoms, as defined in Section \ref{sec:freqPoly}
of the supplement, and same bin width as the MCMC frequency polygon.
The online mean field variational Bayes (MFVB) approximations, corresponding
to Algorithm 5 of Luts \textit{et al.} (2014), are Normal density functions.
The dashed vertical line corresponds to $\betaoTrue=9.36$.
}}
\label{fig:logisRegCompar}
\end{figure}

Suppose that new predictor/response pairs $(\xnew,\ynew)$ are generated 
according to 
\begin{equation}
\xnew\sim\mbox{Uniform}(0,1),\quad
\ynew|\xnew\sim\mbox{Bernoulli}\big(\expit(\betazTrue + \betaoTrue\,\xnew)\big)
\label{eq:onlLogisRegn}
\end{equation}
where $\betazTrue=-7.5$ and $\betaoTrue=9.36$. Of interest here are the posterior 
density functions of the coefficient parameters
$$\pDens(\beta_0|\bxcurr,\bycurr)\quad\mbox{and}\quad \pDens(\beta_1|\bxcurr,\bycurr)$$
where $\bxcurr$ and $\bycurr$ are the current predictor and response vectors as
the data stream in according to 
$$n\longleftarrow n+1,\quad \bxcurr\longleftarrow(\bxcurr,\xnew)\quad\mbox{and}\quad
\bycurr\longleftarrow(\bycurr,\ynew).$$
We warmed up both Algorithm \ref{alg:onlineSMCforGLM} of the present paper
and Algorithm 5 of Luts \textit{et al.} (2014) with a sample size 
of $\nwarm=100$ and terminated at $n=500$. As a ``gold standard'' we 
also obtain the batch Monte Carlo Markov chain fits to each of the 
$n=100,101,\ldots,500$ data sets using the package 
\textsf{rstan} (Guo \textit{et al.}, 2023) within the 
\textsf{R} computing environment (\textsf{R} Core Team, 2024).
A movie in the supplemental material
\footnote{Currently the movie is on the web-page:
\texttt{http://matt-p-wand.net/MOWmovies.html}}
of this article displays and compares the approximations of 
$\pDens(\beta_0|\bxcurr,\bycurr)$ and $\pDens(\beta_1|\bxcurr,\bycurr)$.
Figure \ref{fig:logisRegCompar} shows four frames of the movie for 
$n\in\{200,300,400,500\}$ and the parameter $\beta_1$.
The approximations to the $\pDens(\beta_j|\bxcurr,\bycurr)$ 
based on batch Markov chain Monte Carlo and online sequential Monte Carlo
are displayed using frequency polygons, as described in Section \ref{sec:freqPoly}
of the supplement with bin width rule (\ref{eq:fpBWrule}). The online mean field 
variational Bayes approximations are Normal density functions.

Figure \ref{fig:logisRegCompar} and its extended movie form 
show that online sequential Monte Carlo leads to very good approximation 
of the posterior distributions of $\beta_0$ and $\beta_1$. In contrast,
online mean field variational Bayes provides overly narrow posterior
density function approximations for this example.

\begin{figure}[!h]
\centering
\includegraphics[width=1.0\textwidth]{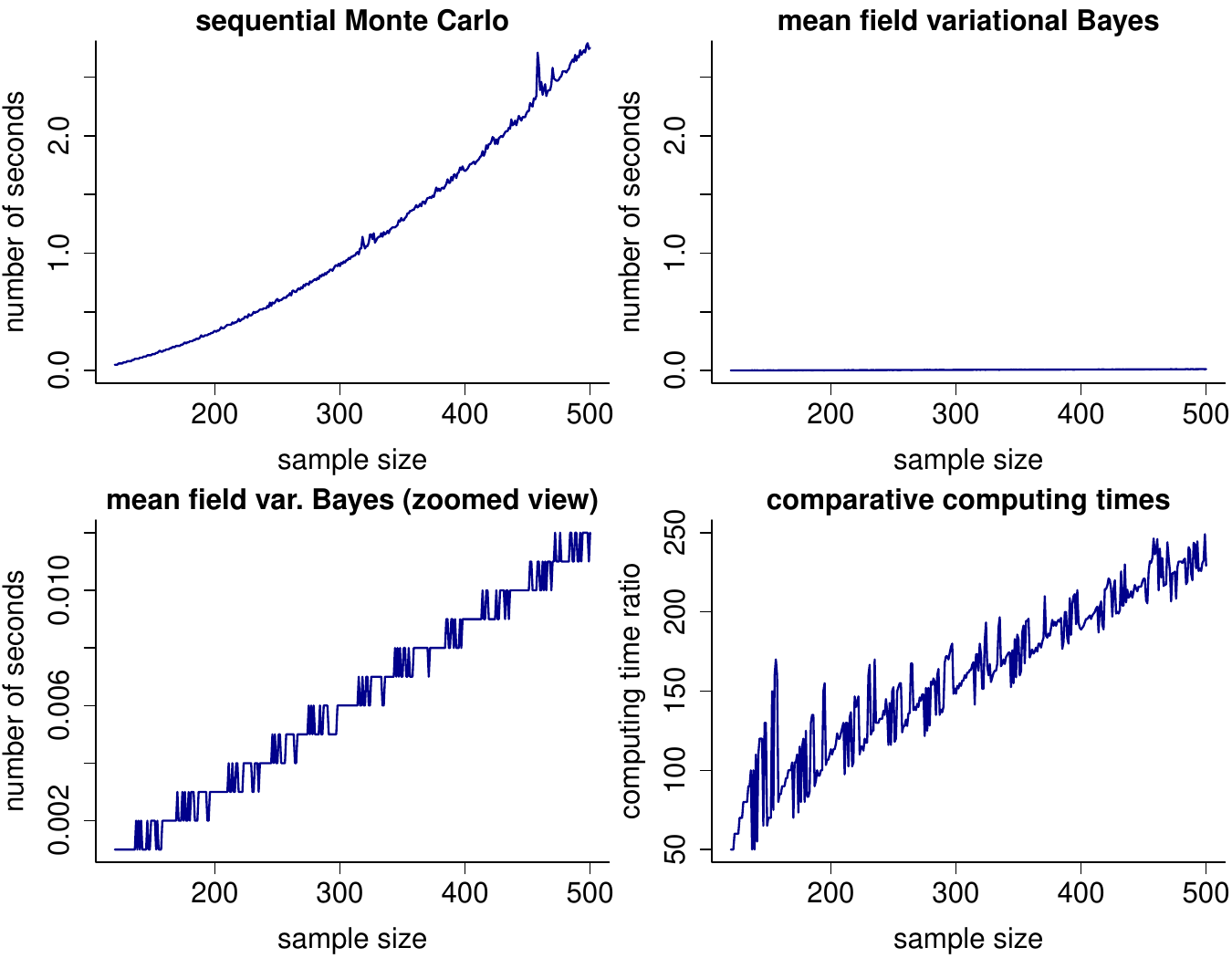}
\caption{\textit{The computing times and their ratios for the
Figure \ref{fig:logisRegCompar} example. In each panel the horizontal
axis corresponds to the sample size in the online fitting phase.
In the upper panels the vertical axis corresponds to number
of seconds required for online fitting and inference after the 
warm-up phase and the axis limits are the same to aid visual comparison.
The lower left panel is a zoomed view of the upper right panel's curve.
The vertical axis in the lower right panel corresponds to 
computing time ratios.}}
\label{fig:SMCvsJJtimes}
\end{figure}

A price to be paid for sequential Monte Carlo's improved accuracy 
is slower computing time. Figure \ref{fig:SMCvsJJtimes} conveys
this cost when running the relevant algorithms in the \textsf{R} 
computing environment (\textsf{R} Core Team, 2024)
on the third author's \textsf{MacBook Air} laptop, which has 
a 3.2 gigahertz processor and 16 gigabytes of random access memory.
The jaggedness in the Figure \ref{fig:SMCvsJJtimes} curves is
due to rounding. It takes the sequential Monte Carlo about 2.75 seconds to get from 
$n=100$ to $n=500$ whereas mean field variational Bayes takes only 
0.01 second. The Figure \ref{fig:SMCvsJJtimes}  curves show the times and their 
ratios for getting from $n=100$ to intermediate sample sizes. The sequential Monte
Carlo computing times appear to be quadratic in sample size 
whilst the mean field variational Bayes
times are linear. The ``lightning fast'' aspect of the 
online mean field variational Bayes needs to be traded off against 
it being prone to inaccurate inference, as Figure \ref{fig:logisRegCompar} 
demonstrates.

\subsection{Online Binary Response Nonparametric Regression}\label{sec:binaryNPRexamp}

This second simulated data illustration involves extension of (\ref{eq:onlLogisRegn}) to 
\begin{equation}
\xnew\sim\mbox{Uniform}(0,1),\quad
\ynew|\xnew\sim\mbox{Bernoulli}\big(\fTrue(\xnew)\big)
\label{eq:binaryNPR}
\end{equation}
for a smooth function $\fTrue$ such that $0\le\fTrue(x)\le1$ for $x\in(0,1)$.
In this section's illustrations we have
$$\fTrue(x)\equiv\{1.05-1.02\,x+0.018\,x^2+0.4\,\phi(x;0.38,0.08)
          +0.08\,\phi(x;0.75,0.03)\}/2.7$$
where $\phi(\cdot;\mu,\sigma)$ denotes the density function of the
$N(\mu,\sigma^2)$ distribution.

Online estimation and inference concerning $\fTrue$ can be achieved using
the $R=1$ version of Algorithm \ref{alg:onlineSMCforGLMM} with the current
$\bX$ and $\bZ$ matrices set to 
\begin{equation}
\bX=\left[
\begin{array}{cc}
1       & x_1    \\[1ex]
\vdots  & \vdots \\
1       & x_n
\end{array}
\right]
\quad\mbox{and}\quad
\bZ=\left[
\begin{array}{ccc}
z_1(x_1)  & \cdots & z_K(x_1)\\[1ex]
\vdots    & \ddots & \vdots  \\[1ex]
z_1(x_n)  & \cdots & z_K(x_n)
\end{array}
\right]
\label{eq:forZOSull}
\end{equation}
where $\{z_k(\cdot):1\le k\le K\}$ is a suitable spline basis such as 
that described in Section 4 of Wand \myand Ormerod (2008).
The $\bC$ matrix appearing in Algorithm \ref{alg:onlineSMCforGLMM} 
is then given by $\bC=[\bX\ \bZ]$. For this example we have $K=37$.

\begin{figure}[!h]
\centering
\includegraphics[width=1.0\textwidth]{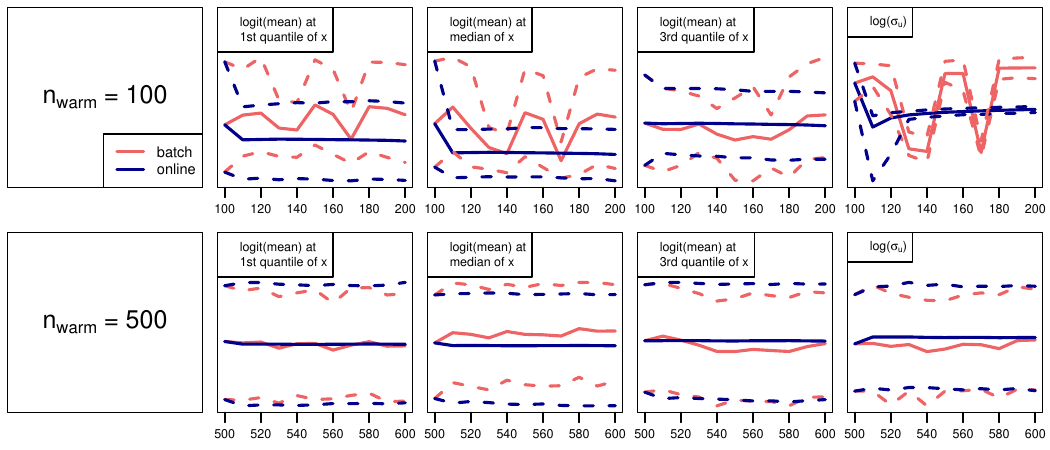}
\caption{\textit{Batch-based convergence diagnostics for the binary response
nonparametric regression example.
The solid lines track posterior means, while the dashed lines show
the limits of corresponding 95\% credible intervals.
First row: the horizontal axes correspond to sample sizes between
a warm-up batch sample of size $\nwarm=100$ and validation samples
up to $\nvalid=100$ greater than $\nwarm$.
Second row: as for the first row, but with $\nwarm=500$.}}
\label{fig:nWarmBinNPR}
\end{figure}

With the online sequential Monte Carlo particles having a dimension of around $40$
we found that a substantial batch-based warm-up was required. This aspect
is conveyed by Figure \ref{fig:nWarmBinNPR} which compares the diagnostic
plots corresponding to the generalized linear mixed model extension
of Algorithm \ref{alg:onlineSMCforLMwTune} for $\nwarm=100$
and $\nwarm=500$. It is apparent from Figure \ref{fig:nWarmBinNPR} 
that the lower warm-up sample size is not adequate and one around five
times larger is desirable for good online estimation and inference
for $\fTrue$.
An analogous phenomenon was observed for the online mean field variational
Bayes approach used by Luts \textit{et al.} (2014), with Figure 5 of that
article showing that $\nwarm=100$ is inadequate for a similar model.
Since the updates occur in a high-dimensional Euclidean space, and the mixing properties
of random walk Metropolis--Hastings algorithms deteriorate in a manner inversely proportional
to the dimension of the distribution being sampled (Gelman, Gilks \myand Roberts, 1997), the
starting values corresponding to lower $\nwarm$ are more susceptible to divergence
away from the correct posterior distributions.

\begin{figure}[!h]
\centering
\includegraphics[width=1.0\textwidth]{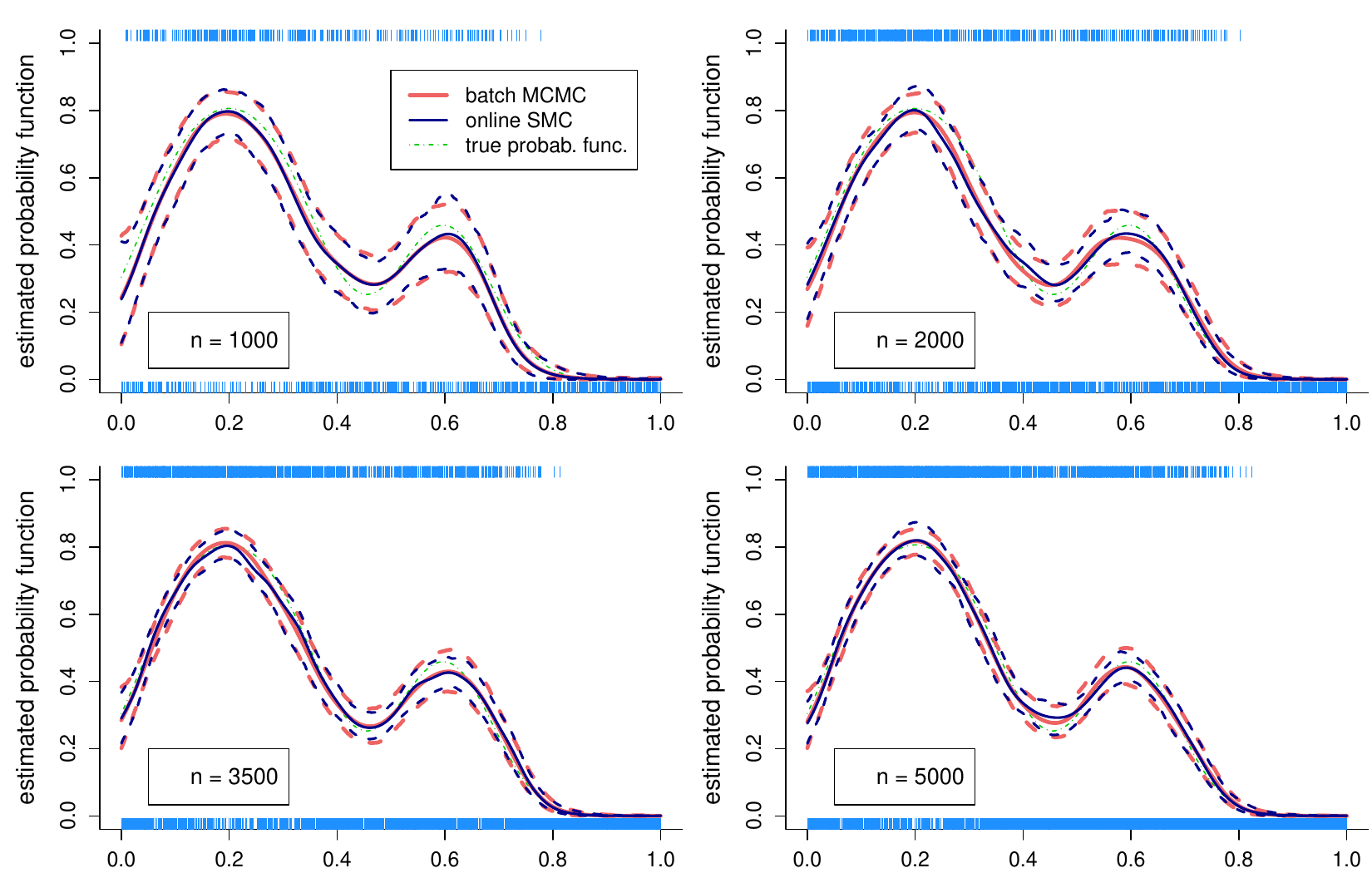}
\caption{\textit{Comparison of online sequential Monte Carlo 
and batch Markov chain Monte Carlo inference for the 
probability function $\fTrue$ in the binary response nonparametric
regression example for four example sample sizes from the movie
in the supplemental material. The solid curves correspond to the
posterior mean, which is targeting the true probability function shown
as a dot-dashed curve. The dashed curves correspond to pointwise
95\% credible intervals.}}
\label{fig:binNPRframes}
\end{figure}

A movie in the supplemental material
\footnote{Currently the movie is on the web-page:
\texttt{http://matt-p-wand.net/MOWmovies.html}}
of this article displays and compares the online sequential
Monte Carlo and batch Markov chain Monte Carlo estimates,
and variability, bands of $\fTrue$. Figure \ref{fig:binNPRframes}
displays four frames of this movie, for the sample sizes 
$n\in\{1000,2000,3500,5000\}$. There is very good correspondence
between the online and batch fits.

\subsection{Illustration for Actual Data}

To illustrate the methodology on actual data we applied Algorithm \ref{alg:onlineSMCforLMM}
for online additive model fitting to sequentially arriving data from
the data frame \texttt{SydneyRealEstate} within the \textsf{R} package 
\textsf{HRW} (Harezlak \textit{et al.}, 2021).
The data consist of numerical attributes of 37,676 houses sold in Sydney, Australia, during 2001.
The response variable is the natural logarithm of sale price in Australian dollars. After conversion
of categorical variables to indicator variables there are around $40$ candidate predictors.
Most of the candidate predictors are continuous and could impact the mean response either 
linearly or non-linearly. Given that this is just an illustration, we first ran the full data
through the generalized additive model selection procedure provided by the
\textsf{R} package \textsf{gamselBayes} (He \myand Wand, 2023) and arrived at 
12 predictors entering the model linearly and 14 predictors entering the model non-linearly. 
Table \ref{tab:SydRealPreds} describes each of these 26 predictors. 
Fuller details are provided in the documentation
of \texttt{SydneyRealEstate} within the \textsf{HRW} package. For the Bayesian model fitting
all variables were linearly transformed to the unit interval and approximate non-informativity
was imposed through the use of the hyperparameter choices $\bmu_{\bbeta}=\bzero$,
$\bSigma_{\bbeta}=10^{10}\bI$ and $\seps=\sur=10^5$. For the upcoming graphical summaries
all posterior distributions were back-transformed to correspond to the original units.
The spline basis functions for each predictor are analogous to 
those given by (\ref{eq:forZOSull}) but with $K=17$.

\begin{table}[ht]
\begin{center}
\begin{tabular}{ll}
\hline
predictors entering linearly  & predictors entering non-linearly \\[0.1ex]
\hline\\[-0.9ex]
degrees longitude                      & lot size\\
distance to nearest highway            & degrees latitude \\
distance to harbour tunnel             & inflation rate measure \\
nitric oxide level                     & average income of suburb\\
suspended matter level                 & distance to nearest bus stop \\
ozone level                            & distance to nearest park\\
particulate matter $<10$ micrometers   & distance to nearest main road \\
sodium dioxide level                   & distance to nearest sealed road \\
distance to nearest medical services   & distance to nearest unsealed road\\
indicator sale in 2nd quarter          & proportion of foreigners in suburb\\
indicator sale in 3rd quarter          & distance to nearest ambulance \\
indicator sale in 4th quarter          & distance to nearest factory \\
                                       & distance to nearest hospital \\
                                       & distance to nearest school\\
\hline
\end{tabular}
\caption{\textit{Predictors used in the online additive model fitting illustration
for real estate data for houses sold in Sydney, Australia, during 2001.}}
\label{tab:SydRealPreds}
\end{center}
\end{table}

\begin{figure}[!h]
\centering
\includegraphics[width=0.94\textwidth]{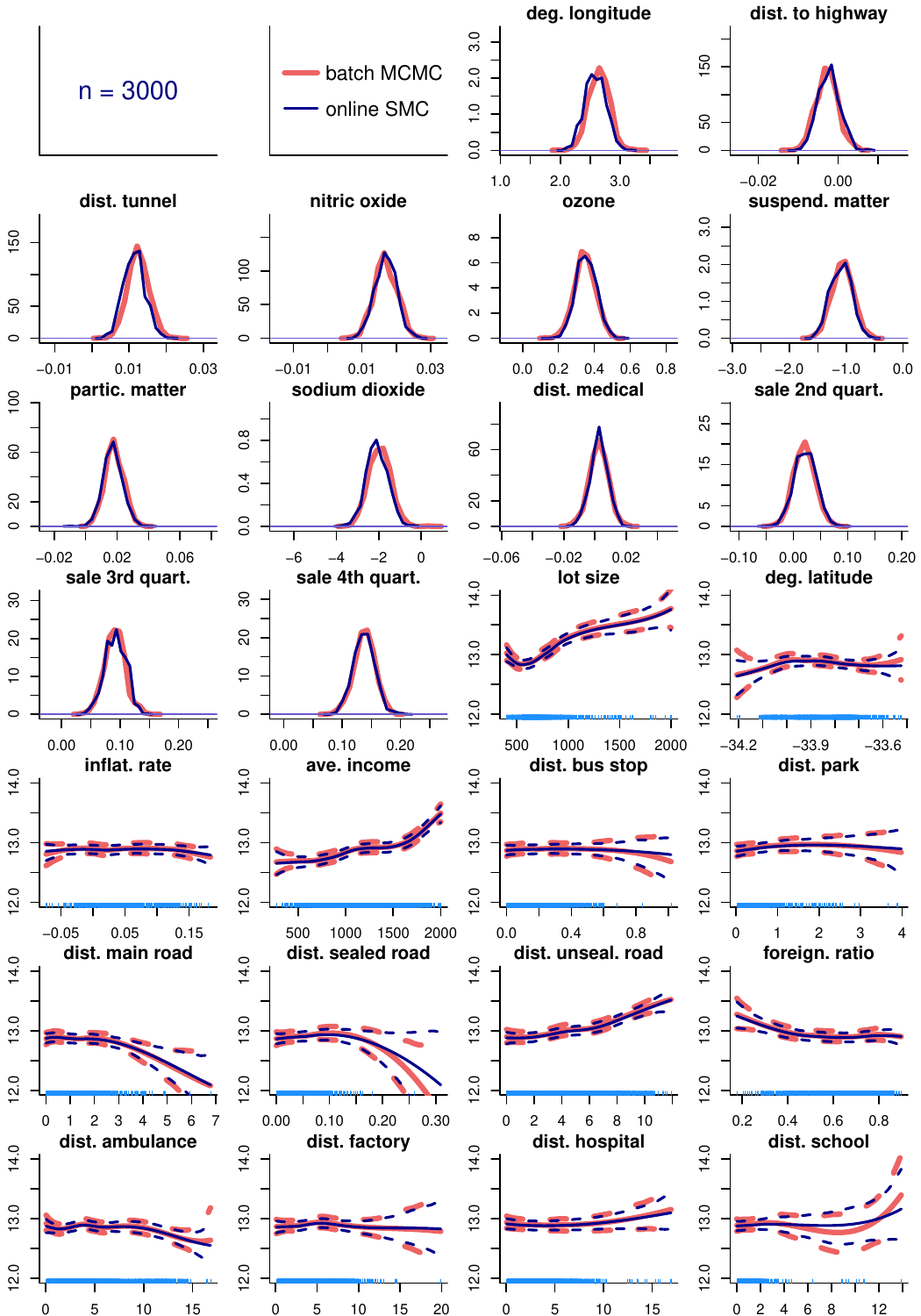}
\caption{\textit{Comparison of online sequential Monte Carlo 
and batch Markov chain Monte Carlo inference for the additive model
fits to the Sydney real estate data for a sample size of $n=3000$.
The frequency polygon plots correspond to approximate 
posterior density functions of the coefficients for predictors 
entering the model linearly. The subsequent panels show the estimates
of the nonlinear effects for predictors entering the model nonlinearly.
The solid curves are posterior means, with each other predictor set 
to its median value. The dashed curves are pointwise 95\% credible intervals.
}}
\label{fig:SydReaEstFrame}
\end{figure}

Algorithm \ref{alg:onlineSMCforLMM} was warmed up with a batch Markov chain Monte Carlo
fit to the first $n=1000$ fields of \texttt{SydneyRealEstate}. This was followed by
online additive model updating for sequentially arriving data based on the
next $4000$. For comparison, we then obtained the batch fits for each of the 
$n\in\{1000,1010,1020,\ldots,5000\}$ data sets. A movie in the supplemental material
\footnote{Currently the movie is on the web-page:
\texttt{http://matt-p-wand.net/MOWmovies.html}}
shows the online additive model fits and compares them with their batch counterparts.
Figure \ref{fig:SydReaEstFrame} shows the $n=3000$ frame of the movie. In both the 
movie and Figure \ref{fig:SydReaEstFrame}, the approximate posterior density functions
of the coefficients for the linear effect predictors are displayed using frequency 
polygons as described in Section \ref{sec:freqPoly} of the supplement. The non-linear effect plots 
correspond to slices of fitted surface with all other predictors set to their
median values. The solid curves show posterior means and the dashed curves show
pointwise 95\% credible intervals. Both the movie and Figure \ref{fig:SydReaEstFrame} 
demonstrate online Bayesian inference via Algorithm \ref{alg:onlineSMCforLMM}  
essentially matching the results from successive batch analyses.

\null\vfill\eject

\section{Concluding Remarks}\label{sec:conclud}

We have demonstrated that sequential Monte Carlo provides a viable approach to online, or real-time,
semiparametric regression that overcome the accuracy shortcomings of the mean field variational
Bayes approach. Our algorithms facilitate straightforward implementation in a wide range of
semiparametric regression settings. For generalized response models and particular applications, 
some modifications concerning the Metropolis-Hastings steps may be worth considering. We have
provided the foundations upon which such modifications could be carried out. Extensions to more elaborate
models can also be entertained, with the current article as a solid basis.

\section{Acknowledgements}

We are grateful to David Leslie and Matt McLean for their contributions to this research. 
We also acknowledge helpful reviewer comments. This work was funded by Australian 
Research Council Discovery Project DP140100441.

\section*{References}

\bib
Bon, J.J., Lee, A. \myand Drovandi, C. (2021). Accelerating sequential Monte Carlo with 
surrogate likelihoods. \textit{Statistics and Computing}, \textbf{31}, 
Article number 62.

\bib
Carroll, R.J. (1976).
On sequential density estimation. 
\textit{Zeitschrift f\"ur Wahrscheinlichkeitstheorie und Verwandte Gebiete},
\textbf{36}, 137--151.

\bib
Chopin, N. \myand Papaspiliopoulos, O. (2020).
\textit{An Introduction to Sequential Monte Carlo}.
Cham, Switzerland: Springer.

\bib
Del Moral, P., Doucet, A. \myand Jasra, A. (2006).
Sequential Monte Carlo samplers.
\textit{Journal of the Royal Statistical Society, Series B},
\textbf{68}, 411--436.

\bib
Fearnhead, P. \myand Taylor, B.M. (2013). An adaptive sequential Monte Carlo sampler. 
\textit{Bayesian Analysis}, \textbf{8}, 411--438.

\bib
Gelman, A., Carlin, J.B., Stern, H.S., Dunson, D.B., Vehtari, A. \myand Rubin, D.B. (2014). 
\textit{Bayesian Data Analysis, Third Edition}, Boca Raton, Florida: CRC Press.

\bib
Gelman, A., Gilks, W., Roberts, G. (1997). Weak convergence and optimal scaling 
of random walk Metropolis algorithms. 
\textit{Annals of Applied Probability}, \textbf{7}, 110–-120.

\bib
Gilks, W.R. \myand Berzuini, C. (2001). Following a moving target -- Monte Carlo inference
for dynamic models. \textit{Journal of the Royal Statistical Society, Series B}, 
\textbf{63}, 127--146.

\bib
Girolami, M. \myand Calderhead, B. (2011). Riemann manifold Langevin and 
Hamiltonian Monte Carlo methods (with discussion).
\textit{Journal of the Royal Statistical Society, Series B},
\textbf{73}, 123--214.

\bib
Gramacy, R.B. \myand Polson, N.G. (2011). Particle learning of Gaussian process models for
sequential design and optimization. \textit{Journal of Computational and Graphical Statistics}, 
\textbf{20}, 102--118.

\bib
Guo, J., Gabry, J., Goodrich, B. \myand Weber, S. (2023). 
\textsf{rstan}: \textsf{R} interface to \textsf{Stan}. 
\textsf{R} package version 2.21.8.
\texttt{https://cran.r-project.org}

\bib
Harezlak, J., Ruppert, D. \myand Wand, M.P. (2018).
{\it Semiparametric Regression with R}. New York: Springer.  

\bib
Harezlak, J., Ruppert, D. \myand Wand, M.P. (2021). \textsf{HRW}.
Datasets, functions and scripts for semiparametric regression supporting 
Harezlak, Ruppert \& Wand (2018).
\textsf{R} package version 1.0. {\tt https://CRAN.R-project.org}

\bib
He, V.X. \myand Wand, M.P. (2023). 
\textsf{gamselBayes}: Bayesian generalized additive model selection. 
\textsf{R} package version 2.0.
\texttt{https://cran.r-project.org}

\bib
Huang, A. \myand Wand, M.P. (2013).
Simple marginally noninformative prior distributions 
for covariance matrices. \textit{Bayesian Analysis}, 
{\bf 8}, 439--452.

\bib
Jaakkola, T.S. \myand Jordan, M.I. (2000). Bayesian parameter estimation
via variational methods. \textit{Statistics and Computing}
\textbf{10}, 25--37.

\bib
Kong, A., Liu, J. \myand Wong, W. (1994).
Sequential imputations and Bayesian missing data problems.
\textit{Journal of the American Statistical Association},
\textbf{89}, 278--288.

\bib
Krzyzak, A. \myand Pawlak, M. (1982). Almost everywhere convergence 
of recursive kernel regression function estimates. 
In \textit{Probability and Statistical Inference:
Proceedings of the 2nd Pannonian Symposium on Mathematical
Statistics}, editors: W. Grossman, G.C. Pflug \myand W. Wertz,
pp. 191--209. 

\bib
Liu, J.S. \myand Chen, R. (1998).
Sequential Monte Carlo methods for dynamic systems.
\textit{Journal of the American Statistical Association},
\textbf{93}, 1032--1044.

\bib
Luo, L \myand Song, P.X.K. (2023).
Multivariate online regression analysis with heterogeneous streaming data.
\textit{Canadian Journal of Statistics}, \textbf{51}, 111--133.

\bib
Luts, J., Broderick, T. \myand Wand, M.P. (2014).
Real-time semiparametric regression.
\textit{Journal of Computational and Graphical Statistics},
{\bf 23}, 589--615.

\bib
Pitt, M. \myand Shephard, N. (1999).
Filtering via simulation: auxiliary particle filters.
\textit{Journal of Computational and Graphical Statistics},
{\bf 94}, 590--599.

\bib
Plummer, M. (2022). \textsf{rjags}: Bayesian graphical models using
Markov chain Monte Carlo. \textsf{R} package version 4-13.
\texttt{https://cran.r-project.org/package=rjags}

\bib
\textsf{R} Core Team (2024). \textsf{R}: A language and environment for statistical
computing. \textsf{R} Foundation for Statistical Computing, Vienna, Austria.
\texttt{https://www.R-project.org/}

\bib
Roberts, G.O. \myand Rosenthal, J.S. (2001).
Optimal scaling for various Metropolis-Hastings algorithms.
{\it Statistical Science}, {\bf 16}, 351--367.

\bib
Roberts, G.O. \myand Stramer, O. (2003).
Langevin diffusions and Metropolis-Hastings algorithms.
{\it Methodology and Computing 
in Applied Probability}, {\bf 4}, 337--358.

\bib
Ruppert, D., Wand, M.P. \myand Carroll, R.J. (2003).
{\it Semiparametric Regression}.
New York: Cambridge University Press.

\bib
Stan Development Team (2022). Stan Modeling Language Users Guide
and Reference Manual. Version 2.30. \texttt{https://mc-stan.org}

\bib
Talagala, P.D. Hyndman, R.J., Smith-Miles, K., Kandanaarachchi, S.
\myand Mu$\tilde{\mbox{n}}$oz, M.A. (2020). Anomaly detection in streaming 
non-stationary temporal data. \textit{Journal of Computational and
Graphical Statistics}, \textbf{29}, 13--27.

\bib
Tierney, L. (1994). Markov chains for exploring posterior distributions.
\textit{The Annals of Statistics}, {\bf 22}, 1701--1728.

\bib
Wand, M.P. \myand Ormerod, J.T. (2008).
On semiparametric regression with O'Sullivan penalized splines.
{\it Australian and New Zealand Journal of Statistics},
{\bf 50}, 179--198.

\bib
Weng, R.C.-H. \myand Coad, D.S. (2018).
Real-time Bayesian parameter estimation for item response models.
\textit{Bayesian Analysis}, \textbf{13}, 115--137.

\bib
Yamato, H. (1971). Sequential estimation of a continuous probability density
function and mode. \textit{Bulletin of Mathematical Statistics}, 
\textbf{14}, 1--12.

\bib
Yin, G.G. \myand Yin, K. (1996). Passive stochastic approximation with
constant step size and window width. \textit{IEEE Transactions on
Automatic Control}, \textbf{41}, 90--106.

\null\vfill\eject
%
%



\renewcommand{\theequation}{S.\arabic{equation}}
\renewcommand{\thesection}{S.\arabic{section}}
\renewcommand{\thetable}{S.\arabic{table}}
\renewcommand{\thefigure}{S.\arabic{figure}}
\setcounter{equation}{0}
\setcounter{table}{0}
\setcounter{figure}{0}
\setcounter{section}{0}
\setcounter{page}{1}
\setcounter{footnote}{0}
\begin{center}
{\Large Supplement for:}
\vskip3mm
\centerline{\Large\bf Online Semiparametric Regression via}
\vskip1.5pt
\centerline{\Large\bf Sequential Monte Carlo}
\vskip4mm
\centerline{\normalsize\sc Marianne Menictas$\null^1$, Chris J. Oates$\null^2$ 
\myand Matt P. Wand$\null^3$}
\vskip4mm
\centerline{\textit{$\null^1$Grubhub Inc., U.S.A.,
$\null^2$Newcastle University, U.K.}}
\vskip0.5pt
\centerline{\textit{and $\null^3$University of Technology Sydney, Australia}}
\end{center}

\section{Sequential Monte Carlo Details}\label{sec:SMCdetails}

We now provide details on the sequential Monte Carlo approach
to online semiparametric regression described
in Figure \ref{fig:genericFormSMC}. Note, however, that
the approach is quite generic and holds for wide classes
of Bayesian graphical models. Our presentation is at the generic level.

Our description builds up to the final algorithm in stages.
We start with two simplistic versions of sequential Monte Carlo
for online fitting which omit the resampling enhancement
provided by the \SystematicResample\ algorithm.
This temporary omission allows the essence of the approach 
to be explained more straightforwardly. 

\subsection{Simplistic Versions of Sequential Monte Carlo}

For each of the semiparametric regression models of interest, the
starting situation of no data observed corresponds to $n=0$. 
The arrival of the first response observation $y_1$, and its
corresponding predictor vector, leads incrementation of the 
sample size to $n=1$. Therefore, in terms of the response data,
we have:

\begin{center}
\begin{tabular}{lcccccc}
sample size:  & $n=0$        & $n=1$  & $n=2$  & $n=3$ & $n=4$ & $\cdots$ \\[0ex]
new response: &              &  $y_1$ & $y_2$  & $y_3$ & $y_4$ & $\cdots$ 
\end{tabular}
\end{center}

\noindent
Each of our semiparametric regression models are such that, given the 
full set of parameters $\btheta$, the responses are conditionally 
independent. Hence, for any $n\in\naturalNumbers$:
\begin{equation}
\pDens(y_1,y_2,\ldots,y_n|\btheta)
=\pDens(y_1|\btheta)\pDens(y_2|\btheta)\cdots\pDens(y_n|\btheta).
\label{eq:Guttorp}
\end{equation}

\subsubsection{Fixed Particle Case}\label{sec:fixPartic}

The fixed particle case corresponds to the situation where
the atoms of the probability mass function approximations
of the posterior density functions
$$\pDens(\btheta|y_1),\ \pDens(\btheta|y_1,y_2),\ \pDens(\btheta|y_1,y_2,y_3),
\ \pDens(\btheta|y_1,y_2,y_3,y_4),\ldots$$
are held fixed as the data stream in. At this point we stress that keeping the
atoms fixed is not recommended in practice. We only do it here with pedagogy
in mind. Denote the atoms by 
\begin{equation}
\btheta_1,\ldots,\btheta_M.
\label{eq:AtomAnt}
\end{equation}

Let $n$ be the current sample size and $\by_n$ be the corresponding vector of response values.
The natural probability vector to put on the atoms is 
$$\bp^{[n]}\equiv\Big(\pDens(\btheta_1|\by_n),\ldots,\pDens(\btheta_M|\by_n)\Big)
\Big/\sum_{m=1}^M\pDens(\btheta_m|\by_n).$$
Suppose that a new response observation $\ynew$ arrives so that the sample size becomes $n+1$
and the response vector becomes $\by_{n+1}\equiv(\by_n,\ynew)$. For the probability mass 
approximation of 
$\pDens(\btheta|\by_{n+1})$ the natural probability vector to put on the atoms is 
$$\bp^{[n+1]}\equiv\Big(\pDens(\btheta_1|\by_{n+1}),\ldots,\pDens(\btheta_M|\by_{n+1})\Big)
\Big/\sum_{m=1}^M\pDens(\btheta_m|\by_{n+1}).
$$
In view of (\ref{eq:Guttorp}), as a function of $m\in\{1,\ldots,M\}$, we have

$$\frac{\bp^{[n+1]}_m}{\bp^{[n]}_m}\propto\frac{\pDens(\btheta_m|\by_{n+1})}
{\pDens(\btheta_m|\by_n)}
=\frac{\pDens(\by_{n+1}|\btheta_m)\pDens(\btheta_m)}
{\pDens(\by_n|\btheta_m)\pDens(\btheta_m)}
=\pDens(\ynew|\btheta_m).
$$
Therefore, if the probability vector is initialized as $\bp_m^{[0]}=1/M$, the
update steps
\begin{equation}
\bp_m\thickarrow\bp_m\pDens(\ynew|\btheta_m)\ \ \ ;\ \ \
\bp_m\thickarrow\bp_m\Big/\sum_{m'=1}^M\bp_{m'}
\label{eq:UnderwoodStreet}
\end{equation}
lead to the required probability mass function approximations as the data stream in.

The directed acyclic graphical nature of semiparametric regression models means that
(\ref{eq:UnderwoodStreet}) often can be reduced to a simpler expression, 
such that $\btheta_m$ is replaced by the particle sub-vectors corresponding to the
parents of $\by_n$. As an example, for the logistic additive model given 
by (\ref{eq:lastGasp}) and with directed acyclic graph representation in 
Figure \ref{fig:logisAddModDAG} the update multiplicative factor simplifies
to $\pDens\big(\ynew|\bbeta_m,(\bu_1)_m,(\bu_2)_m\big)$.

Holding the particles to be fixed throughout the online analysis is far from
desirable. Improved probability mass function approximation would be realised by
having the particles reflect the ever-changing location, spread and shape of 
$\pDens(\btheta|\by_n)$ as $n$ increases. Section \ref{sec:theNext} discusses
one established strategy for varying the particles.

\subsubsection{Varying Particles Adjustment}\label{sec:theNext}

Several strategies have been developed for varying the particles after the arrival
of a new observation vector. The one which we adopt for online semiparametric
regression is known as the \emph{resample-move} algorithm. Its description and 
justification are given in Section 3 of Gilks \myand Berzuini (2001), and is based
on similar ideas from e.g. Kong \textit{et al.} (1994). The algorithm involves 
both \emph{resample} and \emph{move} steps, which we now describe in turn.

The \emph{resample} step involves drawing a sample of size 
$M$ from the probability mass function with atoms corresponding to the
current particles and probabilities corresponding to the current probability 
vector. It is motivated by circumvention of a problem known as \emph{degeneracy},
which is a tendency for probability mass to be confined to a small number of particles
as the iterations progress. The \SystematicResample\ algorithm facilitates 
the resample step, using systematic resampling (see Section \ref{sec:DDSjustif} for details), 
for particles stored in matrix form. However, it is recommended 
that the resample step only be carried out when $\bp^T\bp$ is above a 
particular threshold (e.g. Chopin \myand Papaspiliopoulos, 2020).
We provide the justification for this aspect of the approach in Section \ref{sec:justifSSp}
of the supplement.

The \emph{move} step is based on Markov chain theory, and involves use of 
a transitional kernel with an invariant distribution. Moving the particles
helps ensure that the ever-changing posterior density functions are 
well-approximated by the current particles and their probability vector as new data are observed. 
For the purposes of online semiparametric regression, the move step simply 
corresponds to draws from the current full conditional distributions. For illustration,
consider the logistic additive model given by (\ref{eq:lastGasp}) and with
directed acyclic graph representation in Figure \ref{fig:logisAddModDAG}.
Then the move step is as follows (with $\bCcurr$ defined below):
\begin{itemize}
\item[] For $m=1,\ldots,M$:
\begin{itemize}
\item[] $[\bbeta^T\ \bu_1^T\ \bu_2^T]_m^T\thickarrow\mbox{draw from the distribution having 
density function proportional to}$
\item[] $\begin{array}{l}
\qquad\qquad\qquad\exp\Big[\bycurr^T\bCcurr[\bbeta^T\ \bu_1^T\ \bu_2^T]^T
-\bone^T\log\big\{\bone+\exp\big(\bCcurr[\bbeta^T\ \bu_1^T\ \bu_2^T]^T\big)\big\}\\
\qquad\qquad\qquad\qquad -\smhalf(\bbeta-\bmu_{\bbeta})^T\bSigma_{\bbeta}^{-1}(\bbeta-\bmu_{\bbeta})
-{\displaystyle\frac{\Vert\bu_1\Vert^2}{2\sigma_{u1}^2}}
-{\displaystyle\frac{\Vert\bu_2\Vert^2}{2\sigma_{u2}^2}}
\Big]
\end{array}
$
\item[] $(a_{u1})_m\thickarrow \mbox{draw from Inverse-Gamma
$\big(1,(\sigma_{u1}^2)_m^{-1}+s_{\sigma^2}^{-1}\big)$}$
\item[] $(\sigma_{u1}^2)_m\thickarrow \mbox{draw from Inverse-Gamma
$\big(\smhalf(K_1+1),(a_{u1})_m^{-1}+\smhalf\Vert(\bu_1)_m\Vert^2\big)$}$
\item[] $(a_{u2})_m\thickarrow \mbox{draw from Inverse-Gamma
$\big(1,(\sigma_{u2}^2)_m^{-1}+s_{\sigma^2}^{-1}\big)$}$
\item[] $(\sigma_{u2}^2)_m\thickarrow \mbox{draw from Inverse-Gamma
$\big(\smhalf(K_2+1),(a_{u2})_m^{-1}+\smhalf\Vert(\bu_2)_m\Vert^2\big)$}$
\end{itemize}
\end{itemize}
With the exception of $[\bbeta\ \bu_1\ \bu_2]$, the particle updates are straightforward.
For the $[\bbeta\ \bu_1\ \bu_2]_m$ updates, the matrix $\bCcurr$ is the current design
matrix containing the $x_{1i}$ and $x_{2i}$ predictor data as well as the spline bases
evaluated at these predictors. The required draw for updating $[\bbeta\ \bu_1\ \bu_2]_m$ 
does not involve a standard distribution, and remedies such as Metropolis-Hastings sampling
or slice sampling (e.g. Gelman \textit{et al.}, 2014; Chapters 11--12)
are required to achieve this part of the move step. Algorithm \ref{alg:onlineSMCforGLMM}
treats a more general version of (\ref{eq:lastGasp}) and calls upon 
the random walk proposal version of Metropolis-Hastings sampling for the
$[\bbeta\ \bu_1\ \bu_2]_m$ particle moves.

\subsubsection{Justification of the $\bp^T\bp$ Threshold}\label{sec:justifSSp}

We now explain the justification for checking that the sum of squares of 
the sequential Monte Carlo probability vector $\bp$ is above a particular threshold
and, if it is, applying the \textsc{SystematicSample} algorithm to the particles.

As in Section \ref{sec:discAppx}, let $\theta$ be a generic model parameter of
interest and $\by$ denote the current observed data. Then the atoms of the 
current set of particles 
$$\theta_{1},\ldots,\theta_{M}$$
may be thought of as being a weighted sample from the $\theta|\by$ distribution
with weights stored in $\bp$. The current approximation to the posterior mean of $\theta$
is
$$\sum_{m=1}^M \bp_m\theta_m\quad\mbox{which has conditional variance}\quad
\Var\left(\sum_{m=1}^M \bp_m\theta_m\Big|\by\right)=\bp^T\bp\,\Var(\theta|\by).
$$
The extremal situations are 
$$\Var\left(\sum_{m=1}^M \bp_m\theta_m\Big|\by\right)
=\left\{
\begin{array}{ll}
{\textstyle\frac{1}{M}}\,\Var(\theta|\by) &  \mbox{in the discrete Uniform case}\\[1ex]
\Var(\theta|\by) &  \mbox{in the degenerate case},\\
\end{array}
\right.
$$
where the former case is that where $\bp_m=1/M$ for all $1\le m\le M$ and the
latter case is that where all of the probability mass is on a single particle.
The threshold 
\begin{equation}
\bp\bp^T>\frac{2}{M}=\mbox{twice the minimum possible value of the discrete Uniform case}
\label{eq:HoneyOnCrumpet}
\end{equation}
serves as a reasonable default for avoiding degeneracy.

Lastly, we relate (\ref{eq:HoneyOnCrumpet}) to the \emph{effective sample size}
notion from the importance sampling and sequential Monte Carlo literatures  
(e.g. Chopin \myand Papaspiliopoulos, 2020; Section 8.6). 
Trivially,
$$\bp\bp^T>\frac{2}{M}\quad\mbox{is equivalent to}\quad\mbox{ESS}(\bp)<\frac{M}{2}$$
where
\begin{equation}
\mbox{ESS}(\bp)\equiv 1/\bp^T\bp
\label{eq:CaveFromVic}
\end{equation}
is the most common effective sample size measure. Martino \textit{et al.} (2017), for example, 
provides some alternatives to (\ref{eq:CaveFromVic}).

\section{Algorithm Justifications}

The essence of online semiparametric regression via sequential Monte Carlo
is provided by Algorithms \ref{alg:onlineSMCforLM}--\ref{alg:onlineSMCforGLMM},
each of which depend on the Algorithm \ref{alg:SystematicSample} for the
\textsc{SystematicSample} steps. We now provide full justifications for
them. For completeness, we also justify  Algorithm \ref{alg:batchMCMCforLM},
which uses batch processing.

\subsection{Justification of Algorithm \ref{alg:SystematicSample}}\label{sec:DDSjustif}

Consider a $d$-variate discrete distribution with atoms 
\begin{equation}
\btheta_1,\ldots,\btheta_M\in\real^d\ \mbox{and probability vector}\ \bp\ (M\times1).
\label{eq:PrayToStAnthony}
\end{equation}
Algorithm \ref{alg:SystematicSample} is concerned with obtaining a sample of 
size $M$ from $\{\btheta_1,\ldots,\btheta_M\}$ that is well-behaved 
in terms of combating degeneracy in sequential Monte Carlo schemes.
Systematic resampling  is recommended and widely used within the sequential 
Monte Carlo literature (e.g.\ Chopin \myand Papaspiliopoulos, 2020; Chapter 9).
It involves the following steps:
\begin{itemize}
\item[1.] Let $u$ be a draw from the $\mbox{Uniform}(0,1)$ distribution.
\item[2.] Let $\biota$ be the $(u,u+1,\ldots,u+M-1)/M$ quantiles of the
discrete distribution with atoms $(1,\ldots,M)$ and probability vector $\bp$.
This step involves the quantile function definition given at (\ref{eq:quantileDiscrete}).
\item[3.] The sample from $\{\btheta_1,\ldots,\btheta_M\}$ of size $M$, with replacement, 
corresponds the subscripts in $\biota$.
\end{itemize}

Given $\bp$ and $U$, the $\biota$ vector can be obtained in $O(M)$ steps using e.g.\ the 
cumulative probabilities-based algorithm given in Table 2 of 
Li \textit{et al.} (2015). Such an approach is used in Algorithm \ref{alg:SystematicSample}.

\subsection{Justification of Algorithm \ref{alg:batchMCMCforLM}}\label{sec:batchLMjustif}

Algorithm \ref{alg:batchMCMCforLM} is a standard Gibbs sampling scheme and involves
successive draws from the full conditional distributions of $\bbeta$, $a$ and $\sigma^2$.

Straightforward steps show that full conditional distribution of $\bbeta$ is
$$N(\bOmega^{-1}\bomega,\bOmega^{-1})\quad\mbox{where}\quad\bOmega
\equiv\frac{\bX^T\bX}{\sigma^2}+\bSigma_{\bbeta}^{-1}
\quad\mbox{and}\quad\bomega\equiv \frac{\bX^T\by}{\sigma^2}
+\bSigma_{\bbeta}^{-1}\bmu_{\bbeta}.$$
Therefore, if $\bz\sim N(\bzero,\bI_p)$ then 
\begin{equation}
\bOmega^{-1/2}\bz+\bOmega^{-1}\bomega
\label{eq:SidingSprings}
\end{equation}
is a draw from the full conditional distribution of $\bbeta$. If $\bOmega=\bUOmega\diag(\bdOmega)\bUOmega^T$
is the spectral decomposition of $\bOmega$ then, noting that $\bUOmega^{-1}=\bUOmega^T$, it is easy
to show that (\ref{eq:SidingSprings}) is equal to 
$$\bUOmega\left(\frac{\bUOmega^T\bz}{\sqrt{\bdOmega}}+\frac{\bUOmega^T\bomega}{\bdOmega}\right)$$
which corresponds to the $\bbeta^{[g]}$ update in Algorithm \ref{alg:batchMCMCforLM}. 
This approach has the advantage that, once the spectral decomposition of $\bOmega$ has 
been obtained, no matrix inversion is required for the $\bbeta$ draw.

The draws for the auxiliary variable $a$ and variance parameter $\sigma^2$ involve routine
and simple full conditional derivations.

\subsection{Justification of Algorithm \ref{alg:onlineSMCforLM}}\label{sec:onlineSMCforLMjustif}

First note that Algorithm \ref{alg:onlineSMCforLM} is simply the generic sequential Monte Carlo
algorithm for online semiparametric regression of Figure \ref{fig:genericFormSMC}
applied to the Bayesian multiple linear regression model (\ref{eq:linRegMod})--(\ref{eq:HCtoIG}).

Let $(\bbetaSMC)_m$ be the $m$th column of $\bbetaSMC$ and $(\sigsqSMC)_m$ be the $m$th entry 
of $\sigsqSMC$. The likelihood of $\ynew$ as a function of $(\bbetaSMC)_m$ and $(\sigsqSMC)_m$ 
is
{\setlength\arraycolsep{1pt}
\begin{eqnarray*}
\pDens\big(\ynew\Big|(\bbetaSMC)_m,(\sigsqSMC)_m\big)&=&\big\{2\pi(\sigsqSMC)_m\big\}^{-1/2}
\exp\left[-\frac{\big\{\ynew-(\bbetaSMC)_m^T\bxnew\big\}^2}{2(\sigsqSMC)_m^2}\right]\\[1ex]
&\propto&\exp\Big[\big\{\ynew(\bbetaSMC)_m^T\bxnew-\smhalf\{(\bbetaSMC)_m^T\bxnew\}^2\big\}/(\sigsqSMC)_m\\
&&\qquad-\smhalf\log\big\{(\sigsqSMC)_m\big\}\Big]\\[1ex]
&=&\exp\Big[\Big(\ynew\bdeta_m-\smhalf\bdeta_m^2\Big)/(\sigsqSMC)_m
-\smhalf\log\big\{(\sigsqSMC)_m\big\}\Big]
\end{eqnarray*}
}
where $\bdeta=\bbetaSMC^T\bxnew$.  Therefore, the updates of the 
$1\times M$ vector $\bell$ having $m$th entry equal to 
$\log\big\{\pDens\big(\ynew\Big|(\bbetaSMC)_m,(\sigsqSMC)_m\big)\big\}$
performed by Algorithm \ref{alg:onlineSMCforLM} correspond to the Gaussian
multiple linear regression special case of the first line of (\ref{eq:ForbesCoffee}) and
the $\bell$ vector update is justified. 

\subsection{Justification of Algorithm \ref{alg:onlineSMCforLMwTune}}\label{sec:onlineSMCforLMwTuneJustif}

The batch-based tuning and convergence diagnostics of Algorithm \ref{alg:onlineSMCforLMwTune}
are the Monte Carlo analogues of those conveyed by Algorithm 2' of Luts \textit{et al.} (2014)
for the mean field variational Bayes. In the latter article, such tuning and diagnosis was
found to be important for online semiparametric regression with convergence from the
simple and na\"{\i}ve initializations not always guaranteed.  Similar advice applies to the 
sequential Monte Carlo approach as demonstrated in Section \ref{sec:binaryNPRexamp}.

\subsection{Justification of Algorithm \ref{alg:onlineSMCforLMM}}\label{sec:onlineSMCforLMMjustif}

Algorithm \ref{alg:onlineSMCforLMM} arises from
applying the generic sequential Monte Carlo
algorithm for online semiparametric regression of Figure \ref{fig:genericFormSMC}
to the Gaussian response linear mixed model (\ref{eq:LMM1})--(\ref{eq:LMM3}),
but with the batch-based tuning and convergence diagnosis adjustment 
conveyed by Algorithm \ref{alg:onlineSMCforLMwTune}.

The justification for the $\bell$ vector update is analogous to that
for Algorithm \ref{alg:onlineSMCforLM}. The difference here is that
the coefficients vector includes the $\buSMC$ random effects component.
Also, the $\bxnew$ from Algorithm \ref{alg:onlineSMCforLM} is 
replaced by the $\bcnew$, the new row of the $\bC$ matrix that 
accompanies $\ynew$.

The move steps in the second half of the cycle loop of Algorithm \ref{alg:onlineSMCforLMM}
correspond to draws from the current full conditional distributions of, in order,
the parameters
$$[\bbeta^T\ \bu^T]^T,\ \aeps,\ \sigeps^2,\ a_{u1},\ \sigma^2_{u1},\ldots,a_{uR},\ \sigma^2_{uR}.$$
The derivations of the Gibbsian full conditional distributions involve routine probability
calculus steps. 

The steps involving the $\iStt$, $\iEnd$ and $\bomega$ quantities 
warrant some explanations. Note that the conditional distribution of the $\bu$ 
vector in (\ref{eq:LMM1}) can be rewritten as 
$$\bu_r|\,\sigma_{ur}^2\simind N(\bzero,\sigma_{ur}^2\,\bI_{K_r}),\quad 1\le r\le R,$$
where 
$$[\bu_1^T\cdots\bu_R^T]^T\quad\mbox{is the partition of $\bu$ such that each $\bu_r$ is $K_r\times1$.}$$
For each $1\le r\le R$,  the full conditional distribution of $\sigma_{ur}^2$ is
\begin{equation}
\mbox{Inverse-Gamma}\big(\smhalf(K_r+1),a_{ur}^{-1}+\smhalf\Vert\bu_r\Vert^2\big)
\label{eq:painStake}
\end{equation}
The $\iStt$ and $\iEnd$ updates ensure that the correct row-wise sub-blocks of $\buSMC$
are used to extract the particles that correspond to the $\bu_r$ sub-vectors of $\bu$ 
($1\le r\le R$). For each $1\le m\le M$ and $1\le r\le R$, these particles are stored 
in the temporary vector $\bomega$ and then used for the move step corresponding
to the (\ref{eq:painStake}) full conditional distribution.

\subsection{Justification of Algorithm \ref{alg:onlineSMCforGLM}}\label{sec:NeddieSeagoon}

Algorithm \ref{alg:onlineSMCforGLM} arises from
applying the generic sequential Monte Carlo
algorithm for online semiparametric regression of Figure \ref{fig:genericFormSMC}
to the generalized linear mixed model (\ref{eq:GLMM1}),
but with the batch-based tuning and convergence diagnosis adjustment 
conveyed by Algorithm \ref{alg:onlineSMCforLMwTune}.

If $(\bbetaSMC)_m$ denotes the $m$th column of $\bbetaSMC$ then the likelihood of $\ynew$ 
as a function of $(\bbetaSMC)_m$ is
$$
\pDens\big(\ynew\Big|(\bbetaSMC)_m\big)=
\exp\left\{\ynew(\bbetaSMC)_m^T\bxnew-b\big((\bbetaSMC)_m^T\bxnew\big)+c(\ynew)\right\}
\propto\exp\Big\{\ynew\bdeta_m-b(\bdeta_m\big)\Big\}
$$
where $\bdeta=\bbetaSMC^T\bxnew$.  Hence, the updates to the vector
$1\times M$ vector $\bell$ having $m$th entry equal to 
$\log\big\{\pDens\big(\ynew\big|(\bbetaSMC)_m\big)\big\}$
performed by Algorithm \ref{alg:onlineSMCforGLM} correspond to the
generalized linear model special case of the first line of (\ref{eq:ForbesCoffee}).
This justifies the $\bell$ vector updates.

The move step for the $\bbetaSMC$ particles requires draws from the current full
conditional distribution of $\bbeta$, which has density function
\begin{equation}
\pDens(\bbeta|\bycurr)\propto
\exp\Big\{\bycurr^T\bXcurr\bbeta-\bone^Tb(\bXcurr\bbeta)-\smhalf(\bbeta-\bmu_{\bbeta})^T
\bSigma_{\bbeta}^{-1}(\bbeta-\bmu_{\bbeta})\Big\}.
\label{eq:WhiteLotusBabe}
\end{equation}
where $\bycurr$ and $\bXcurr$ are, respectively, the current response vector
and design matrix. Since (\ref{eq:WhiteLotusBabe}) is not a standard distribution
from which draws can be easily made, the particle moves require a non-Gibbsian
strategy. Algorithm \ref{alg:onlineSMCforGLM} uses a random walk Metropolis-Hastings
approach. The random walk step for possibly updating the $m$th column of $\bbetaSMC$
is
$$\bbetaRW\thickarrow\mbox{draw from 
$N\left((\bbetaSMC)_m,{\displaystyle\frac{\upsilon^2}{n}}\bI\right)$}
\quad\mbox{for some $\upsilon>0$}.$$
The $\upsilon^2/n$ factor helps ensure that the posterior variances of 
the $\bbetaSMC$ entries have the familiar $O_P(1/n)$ asymptotic behaviour.
Then compute and assign the Metropolis-Hastings ratio
$$\alpha\thickarrow\frac{\pDens(\bbetaRW|\bycurr)}{\pDens((\bbetaSMC)_m|\bycurr)}
\quad\mbox{as well as}\quad
u\thickarrow\mbox{draw from Uniform$(0,1)$}.
$$
If $\alpha\ge u$ then  the assignment $(\bbetaSMC)_m\thickarrow\bbetaRW$ should
be made. Otherwise the $m$th column of $\bbetaSMC$ is not updated.
These steps provide the required full conditional draw according to  
Metropolis-Hastings theory (e.g. Gelman \textit{et al.}, 2014; Chapters 11--12)
The $\bbetaSMC$ updates in Algorithm \ref{alg:onlineSMCforGLM} are 
an operationalization of this random walk Metropolis-Hastings approach with
the calculations done on the logarithmic scale, which helps avoid  
underflow/overflow problems.

\subsection{Justification of Algorithm \ref{alg:onlineSMCforGLMM}}\label{sec:onlineSMCforGLMMjustif}

Algorithm \ref{alg:onlineSMCforGLMM} can be justified using arguments
already given in the justifications of Algorithms \ref{alg:onlineSMCforLMM}
and \ref{alg:onlineSMCforGLM}. The $[\bbetaSMC^T\ \buSMC^T]^T$ updates of 
Algorithm \ref{alg:onlineSMCforGLMM} are analogous to the $\bbetaSMC$ updates 
of Algorithm \ref{alg:onlineSMCforGLM}. The $\bsigsqUSMC$ updates are identical 
to those of Algorithm \ref{alg:onlineSMCforLMM}.

\subsection{Frequency Polygonal Visualization of Posterior Distributions}\label{sec:freqPoly}

As in Section \ref{sec:discAppx} we $\theta$ denote a generic univariate parameter in
a statistical model that takes values over a continuum and let $\bycurr$ denote
the currently observed data. The most intuitive way to  visualize the posterior distribution 
of $\theta$ is via a plot of the posterior density function $\pDens(\theta|\bycurr)$.
In this subsection we describe an effective approach to visualizing and comparing 
batch Markov chain Monte Carlo and online sequential Monte Carlo 
approximations of $\pDens(\theta|\bycurr)$ based on the classical 
\emph{frequency polygon} estimator (e.g. Scott, 1985).

In the case of batch Markov chain Monte Carlo, visualization of $\pDens(\theta|\bycurr)$ 
typically involves applying a probability density estimation technique, such as kernel 
density estimation, to the kept sample $\theta^{[1]},\ldots,\theta^{[\Nkept]}$. 
An alternative approach, which is more amenable to comparison with online sequential 
Monte Carlo approximations of $\pDens(\theta|\bycurr)$, is to obtain a histogram from 
the $\theta^{[1]},\ldots,\theta^{[\Nkept]}$ values and then form the corresponding frequency 
polygon by connecting with straight lines the mid-bin heights of the histogram bars. 

For the online sequential Monte Carlo situation, $\pDens(\theta|\bycurr)$ is approximated
by a probability mass function with atom vector $\batheta$ and probability vector $\bptheta$.
If the number of atoms is high, as is the case for online sequential Monte Carlo,
then visualization a probability mass function is challenging. However, 
the frequency polygonal idea has a natural extension to the discrete case, which 
lends itself to simple display of an online sequential Monte Carlo approximation
$\pDens(\theta|\bycurr)$ and comparison with the batch Markov chain Monte Carlo
counterpart. 

Let $\ba$ be a generic vector of atoms in $\real$ and $\bp$ be the corresponding
probability vector. Suppose that $\real$ is partitioned into equi-sized bins with
width $h$. For each $x\in\real$ define the following function of $x$:
\begin{equation}
\frac{\mbox{sum of entries of $\bp$ corresponding to the entries of $\ba$ that are
in the bin containing $x$}}{h}.
\label{eq:MavisBramston}
\end{equation}
The function of $x$ defined by (\ref{eq:MavisBramston}) is piecewise constant over
the bins and, therefore, ``histogram-like''. The frequency polygon representation
of the probability mass function with atoms $\ba$ and probabilities $\bp$ is formed
by connecting with straight lines the mid-bin heights of the function defined by 
(\ref{eq:MavisBramston}). Frequency polygonal representation of the online
sequential Monte Carlo approximations of $\pDens(\theta|\bycurr)$ are obtained
by applying the principle described in this paragraph to the current
$\batheta$ and $\bptheta$ vectors.

Next we address the practical problem of frequency polygon bin width choice, starting with 
that for estimation of $\pDens(\theta|\bycurr)$ based on a batch Monte Carlo Markov chain 
sample of size $\Nkept$. From (2.4) of Scott (1985), the leading term behaviour
of the mean integrated squared error optimal bin width, which we denote by $\hFP^*$, 
is given by
$$\hFP^*=\left[\frac{480}{49\infint \{\pDens''(\theta|\bycurr)\}^2\,d\theta}\right]^{1/5}
\Nkept^{-1/5}\{1+o(1)\}.
$$
The only unknown in the leading term of $h^*$ is the integrated squared second derivative 
of $\pDens(\theta|\bycurr)$, which is also the only unknown for the optimal bandwidth
for kernel density estimation of the same density function -- see, for example, (2.13) of 
Wand \myand Jones (1995). In the special case of kernel density estimation with the Gaussian
kernel $K(x)=(2\pi)^{-1/2}\exp(-x^2/2)$ we have 
$\hFP^*=\big(1280\sqrt{\pi}/49\big)^{1/5}\,\hKDE^*\{1+o(1)\}$ where $\hKDE$ is
the mean integrated squared error optimal bandwidth for kernel density estimation
based on the Gaussian kernel. This leads to the practical frequency polygon bin width 
rule
\begin{equation}
\hhatFP=\left(\frac{1280\sqrt{\pi}}{49}\right)^{1/5}\,\hhatKDE
\label{eq:fpBWrule}
\end{equation}
where $\hhatKDE$ is a consistent estimator of $\hKDE^*$. This can be obtained from
any statistical software package that supports automatic kernel density estimation.
For example, in  the \textsf{R} language the following code leads to a practical and
consistent frequency polygon estimated optimal bin width value based on the Markov chain 
Monte Carlo sample stored in the array \texttt{theta}:

\begin{center}
\begin{verbatim}
library(KernSmooth) ; hHatFP <- (1280*sqrt(pi)/49)^(1/5)*dpik(theta)
\end{verbatim}
\end{center}

\noindent
Here \texttt{dpik()} is a function for automatic bandwidth selection 
within the \textsf{R} package \textsf{KernSmooth} (Wand \myand Ripley, 2023),
and involves consistent estimation of $\infint \{\pDens''(\theta|\bycurr)\}^2\,d\theta$.

The choice of bin width for frequency polygonal representation of the 
online sequential Monte Carlo probability mass functions remains an open problem.
In the comparisons with batch Markov chain Monte Carlo in Figure
\ref{fig:logisRegCompar} and corresponding movie in the supplemental material
we use the same bin widths for each frequency polygon.

\section*{References}

\bib
Li, T., Boli{\'c}, M. \myand Djuri{\'c}, P.M. (2015).
Resampling methods for particle filtering.
\textit{IEEE Signal Processing Magazine}, 
\textbf{32(3)}, 70--86.

\bib
Martino, L., Elvira, V. \myand Louzada, F. (2017).
Effective sample size for importance sampling based on discrepancy measures.
\textit{Signal Processing}, \textbf{131}, 386--401.

\bib
Scott, D.W. (1985).
Frequency polygons: theory and application.
\textit{Journal of the American Statistical Association},
\textbf{80}, 348--354.

\bib
Wand, M. P. and Jones, M. C. (1995). {\it Kernel Smoothing}.
London: Chapman and Hall.

\bib
Wand, M.P. and Ripley, B.D. (2023).
KernSmooth 2.23. Functions for kernel smoothing
corresponding to the book: Wand, M.P.
and Jones, M.C. (1995) "Kernel Smoothing". \textsf{R} package.
{\tt https://CRAN.R-project.org/package=KernSmooth}

\end{document}